\documentclass[fleqn,usenatbib]{mnras}

\usepackage[T1]{fontenc}
\usepackage{ae,aecompl}

\usepackage[normalem]{ulem}
\usepackage{amsmath}
\usepackage{graphicx} 
\usepackage{lscape}
\usepackage{indentfirst}
\usepackage{enumitem}
\usepackage{xspace}

\usepackage{amssymb}
\usepackage[dvipsnames]{xcolor}
\usepackage{url}
\usepackage[flushleft]{threeparttable}

\newcommand {\bc}{\begin {center}}
\newcommand {\ec}{\end {center}}
\newcommand {\be}{\begin {equation}}
\newcommand {\ee}{\end {equation}}
\newcommand {\beq}{\begin {eqnarray}}
\newcommand {\eeq}{\end {eqnarray}}
\newcommand {\comment}[1]{}

\newcommand {\ergs}{{\rm erg\ \rm s^{-1}}}

\newcommand{\ixpe}{{\it IXPE}\xspace}

\newcommand {\gro}{\mbox{GRO~J1008$-$57}\xspace}
\newcommand {\xper}{\mbox{X~Persei}\xspace}

\title[X-ray polarimetry of \xper]
{X-ray polarimetry of X-ray pulsar \xper: 
another orthogonal rotator?}

\author[A. A.~Mushtukov et al.] 
{A.~A.~Mushtukov,$^{1,2}$\thanks{E-mail: alexander.mushtukov@physics.ox.ac.uk}
S.~S.~Tsygankov,$^{3}$
J.~Poutanen,$^{3}$
V.~Doroshenko,$^{4}$
A.~Salganik,$^{5,6}$
E.~Costa,$^{7}$ 
\newauthor
A.~Di Marco,$^{7}$
J.~Heyl,$^{8}$
F.~La Monaca,$^{7}$
% ...
\\
\textit{\small \noindent Affiliations are listed at the end of the paper} 
}

\pubyear{2023}

\begin{document}
\label{firstpage}
\pagerange{\pageref{firstpage}--\pageref{lastpage}}
\maketitle


\begin{abstract} 
\xper is a persistent low-luminosity X-ray pulsar of period of $\sim$835\,s in a Be binary system.
The field strength at the neutron star surface is not known precisely, but indirect signs indicate a magnetic field above $10^{13}$\,G, which makes the object one of the most magnetized known X-ray pulsars.
Here we present the results of observations \xper performed with the \textit{Imaging X-ray Polarimetry Explorer} (\ixpe).  
The X-ray polarization signal was found to be strongly dependent on the spin phase of the pulsar.
The energy-averaged polarization degree in 3--8 keV band varied from several to $\sim$20 per cent over the pulse with a positive correlation with the pulsed X-ray flux. 
The polarization angle shows significant variation and makes two complete revolutions during the pulse period resulting in nearly nil pulse-phase averaged polarization. 
Applying the rotating vector model to the \ixpe data we obtain the estimates for the rotation axis inclination and its position angle on the sky as well as for the magnetic obliquity.
The derived inclination is close to the orbital inclination reported earlier for \xper. 
The polarimetric data imply a large angle between the rotation and magnetic dipole axes, which is similar to the result reported recently for the X-ray pulsar \gro.  
After eliminating the effect of polarization angle rotation over the pulsar phase using the best-fitting rotating vector model, the strong dependence of the polarization degree with energy was discovered with its value increasing from 0\% at $\sim$2 keV to 30\% at 8 keV.
\end{abstract}

\begin{keywords}
magnetic fields -- polarization -- pulsars: individual: X Persei -- stars: neutron -- stars: oscillations -- X-rays: binaries  
\end{keywords}

\section{Introduction}
\label{sec:Intro}

Accretion of matter onto spinning and strongly magnetized neutron stars (NSs) in close binary systems results in a phenomenon of X-ray pulsars (XRPs, see \citealt{2022arXiv220414185M} for review). 
The magnetic field strength at the NS surface in XRPs is typically measured to be $\sim 10^{12}\,{\rm G}$ and in some sources is confirmed to be as high as $\sim 10^{13}\,{\rm G}$ \citep{2019A&A...622A..61S,2022ApJ...933L...3K}.
Such a strong magnetic field affects both geometry of the accretion flow in XRPs and the physical processes of interaction between radiation and matter  \citep{2006RPPh...69.2631H,2022MNRAS.517.4022S}, making XRPs unique labs to study physics under extreme conditions.

\begin{figure*}
\centering
\includegraphics[width=0.97\linewidth]{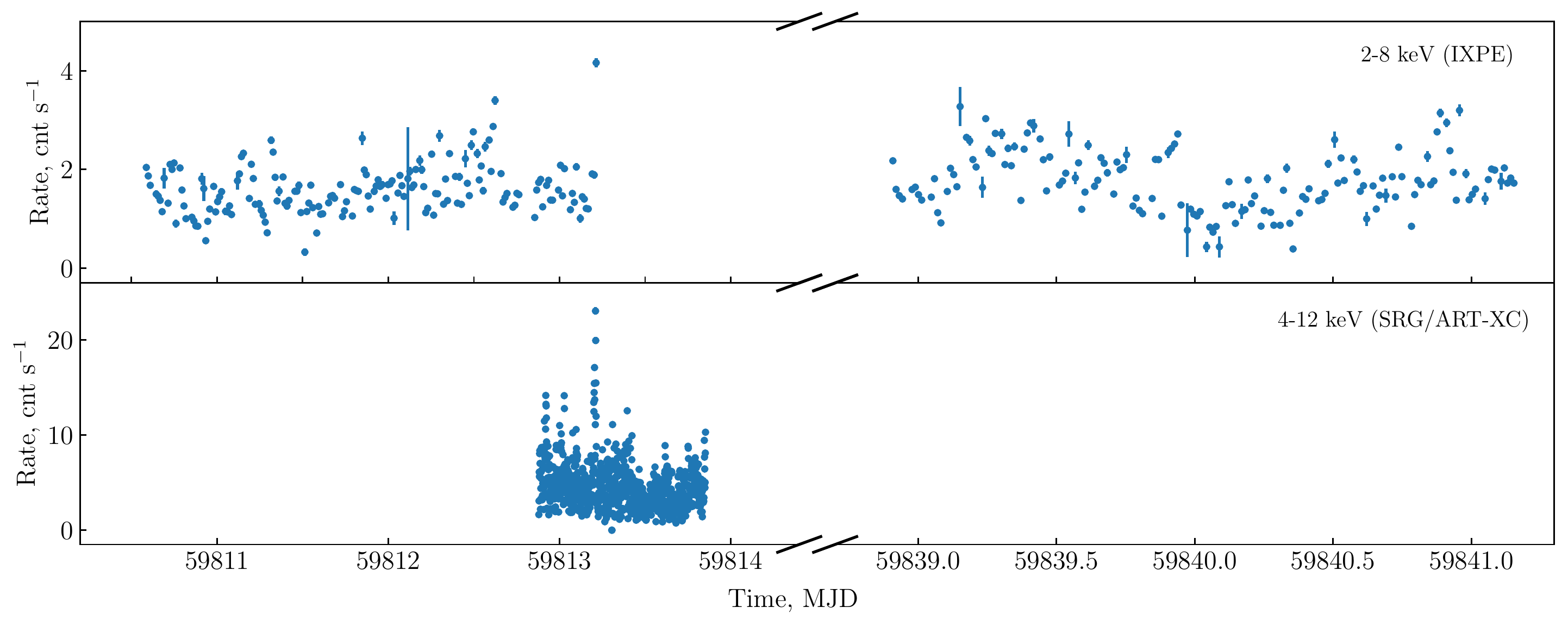}
\caption{{\it Top:} The light curve of \xper in the 2--8 keV energy band summed over three modules of \ixpe. Both observations performed in Aug and Sep 2022 are shown. {\it Bottom:} The light curve of the source in the 4--12 keV band obtained during the quasi-simultaneous {\it SRG}/ART-XC observation.}
 \label{fig:ixpe-lc}
\end{figure*}

{
The geometry of the emitting regions in XRPs is known to be dependent on the mass accretion rate \citep{1976MNRAS.175..395B}: 
at relatively low mass accretion rates ($\dot{M}\lesssim 10^{17}\,{\rm g\,s^{-1}}$), the accretion flow is decelerated in the atmosphere of a NS \citep{1969SvA....13..175Z}, while a high mass accretion rate ($\dot{M}\gtrsim 10^{17}\,{\rm g\,s^{-1}}$) results in the appearance of a radiation dominated shock above the NS surface and extended accretion columns confined by a strong magnetic field and supported vertically by the radiation pressure \citep{1976MNRAS.175..395B,1981A&A....93..255W}.
}

The influence of the strong magnetic field on the radiative transfer in the atmospheres of NSs has long been discussed in the literature. 
In particular, it is well known that the transfer of radiation critically depends on the polarization of X-ray photons and the direction of their propagation with respect to the local direction of magnetic field (see \citealt{1974JETP...38..903G,1983Ap&SS..91..167K} and Chapter 6 in \citealt{1992herm.book.....M} for review). 
Most of the existing models \citep{1985ApJ...298..147M,1985ApJ...299..138M,2021A&A...651A..12S,Caiazzo21} predict a polarization degree up to 80\% at energies below the cyclotron resonance (i.e., at $E<E_{\rm cyc}\approx 11.6\,B_{12}\,{\rm keV}$, where $B_{12}=B/10^{12}\,{\rm G}$ is the local magnetic field strength).
This makes XRPs one of the main targets for the new generation of X-ray polarimeters.

{
Thanks to the successful launch of the first highly sensitive space X-ray polarimeter, the {\it Imaging X-ray Polarimeter Explorer} \citep[\ixpe,][]{Weisskopf2022}, on 2021 December 9, polarization in X-ray energy band has been discovered in XRPs Her~X-1 \citep{2022NatAs...6.1433D}, Cen~X-3 \citep{2022ApJ...941L..14T}, \gro\ \citep{Tsygankov23}, and Vela~X-1 \citep{Forsblom23}.
Detection of polarization in the X-ray energy band allowed to determine the geometrical parameters of the NS, and obtain constraints of the magnetic field structure and the structure of the NS atmosphere.  
Unexpectedly, the observed degree of linear polarization (below 10--15 per cent, even in the phase-resolved data) turned out to be much smaller than that predicted by most theoretical models.
To understand the nature of this discrepancy, we need to observe sources with very different parameters determining geometrical and physical conditions of the emission region.
}

\xper (4U~0352+309) belongs to the rare class of persistent low-luminosity XRPs with Be optical companions \citep[e.g.][]{1999MNRAS.306..100R}. 
Pulsations of the flux with a period of $835\,{\rm s}$ were discovered from the source with the {\it Copernicus} satellite \citep{1976MNRAS.176..201W}. 
The pulsar moves around its optical companion, star HD24534, along a moderately eccentric ($e=0.11\pm0.02$) wide orbit with a period of $P_{\rm orb}=250.3\pm0.6$~d, orbital inclination $i_{\rm orb}\approx30\degr$, a projected semi-major axis of the NS of $a_{\rm x} {\rm sin} i = 454\pm4$~lt-s, and the mass function $f(M)=1.61\pm0.05M_\odot$ \citep{2001ApJ...546..455D}.  
Additional quasi-periodic variations of the observed flux from the source with a period of about 7 years were reported by \citet{2019IAUS..346..131N}.
Recently, the most accurate distance to \xper of $600\pm13$~pc  was determined  from the {\it Gaia} Early Data Release 3 \citep{2021AJ....161..147B}.
The magnetic field of the NS in \xper is not known precisely, but all estimates including broad-band X-ray spectral analyses and the observed evolution of spin frequency point to high magnetic field values.  In particular, a wide depression in the source spectrum observed at around $30\,{\rm keV}$ was interpreted by \citet{2001ApJ...552..738C} and \citet{2012MNRAS.423.1978L} as a cyclotron resonant scattering feature, equivalent to a magnetic field strength of $\sim2.5\times10^{12}$~G. 
However, this spectral feature was later proposed to be related to the double-hump continuum shape typical to XRPs at extremely low mass accretion rates \citep{2019MNRAS.483L.144T}. This would shift the magnetic field to even higher values: $\gtrsim 10^{13}\,{\rm G}$ \citep[e.g.][]{1998ApJ...509..897D,2019MNRAS.487L..30T,2021MNRAS.503.5193M,2021A&A...651A..12S}. 
Applying the accretion torque models to the spin frequency evolution of the source, another estimate of the magnetic field strength can be obtained: $B\sim 4\times10^{13} - 2.5\times10^{14}\,{\rm G}$ \citep{2012A&A...540L...1D,2018PASJ...70...89Y}, which supports the estimate made from the source spectral shape.
{
Combination of a strong magnetic field and low mass accretion rate in \xper point to the accretion from cold disc composed of recombined material (see, e.g., \citealt{2017A&A...608A..17T}).
\xper is the closest low luminosity ($L\sim 10^{35}\,\ergs$) pulsar known, which provides a unique opportunity to study accretion at low rates with good counting statistics essential for polarimetric observations. 
Furthermore, the low mass accretion rate in this object implies that it must have a simple geometrical configuration of the emitting regions at the NS surface, so called hot spots, which simplifies interpretation of the results.}
The magnetic field of X Persei is strong even by the standards of XRPs and makes the source particularly interesting for polarimetric observations as the energy range covered by \ixpe is much below the cyclotron energy, which again simplifies theoretical interpretation of the results.

Here we present the results of \ixpe observations of X Persei performed in August and September 2022.
In Section~\ref{sec:Data}, we present the \ixpe data as well as the accompanying observations by \textit{SRG}/ART-XC. 
Section~\ref{sec:Results} is devoted to the results of the X-ray polarimetric observations. 
We discuss the structure of the atmosphere and the constraints obtained on the geometry in Section~\ref{sec:disc}, and we summarize our findings in Section~\ref{sec:sum}.

\section{Data}
\label{sec:Data}

\begin{figure}
\centering
\includegraphics[width=0.95\linewidth]{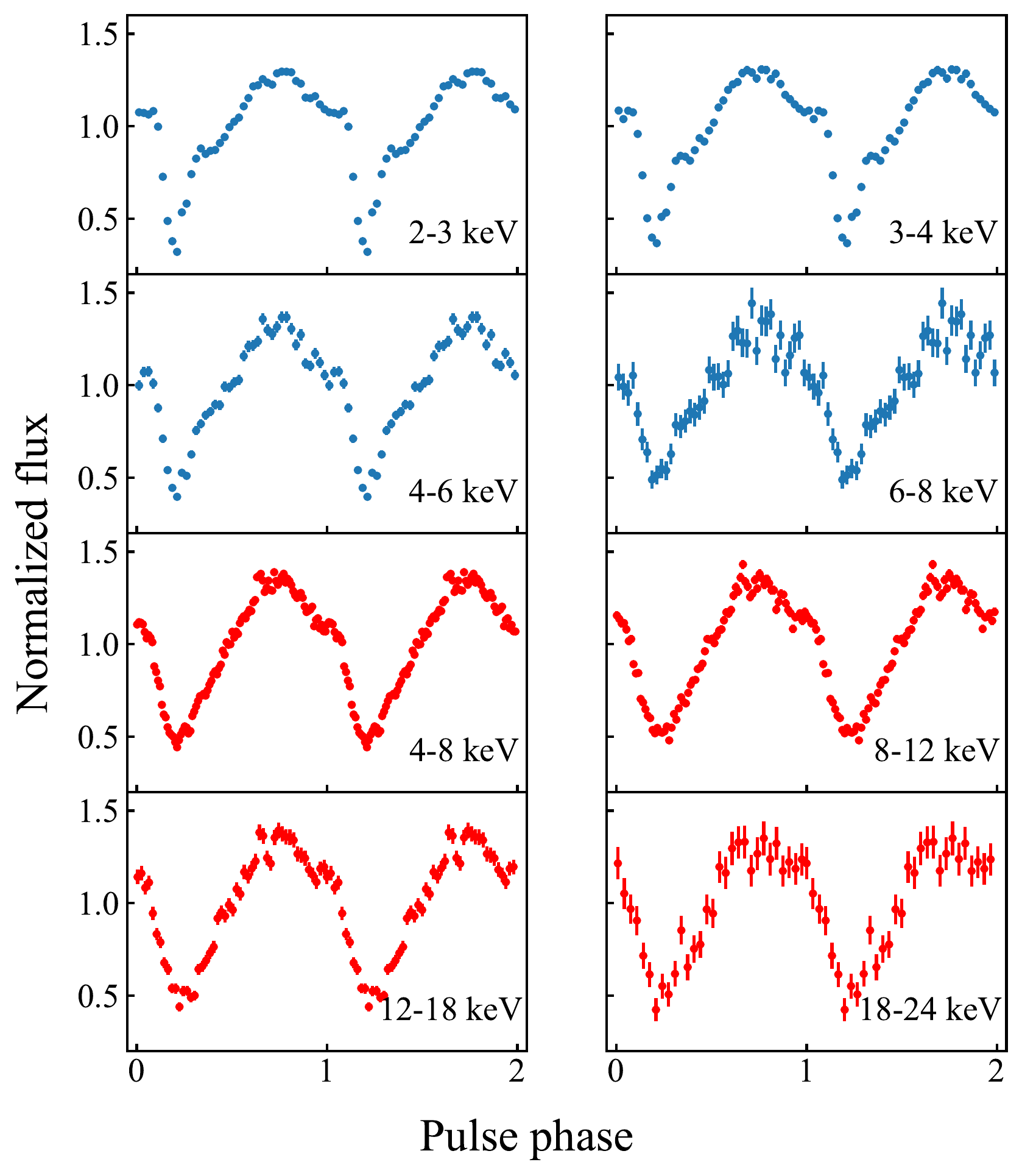}
\caption{Pulse profile of \xper in different energy bands as seen by \ixpe (upper four panels in blue) and ART-XC telescope (bottom four panels in red). 
Data from the three and seven \ixpe and ART-XC units, respectively, were combined.
}
 \label{fig:pprof}
\end{figure}

\subsection{\ixpe}

\xper was observed with the Imaging X-ray Polarimetry Explorer (\ixpe) over the periods of 2022 Aug 19--22 and Sep 16--19 with a total effective exposure of $\simeq$225~ks. 
\ixpe is a joint effort of NASA and the Italian Space Agency, launched by a Falcon 9 rocket on 2021 December 9. The observatory consists of three grazing incidence telescopes, each equipped with an X-ray mirror assembly and a polarization-sensitive detector unit (DU) \citep{2021AJ....162..208S,2021APh...13302628B}. It provides imaging polarimetry over a nominal 2--8 keV band. 
The time resolution and accuracy is $<$10~$\mu$s, far better than what is needed for our analysis. 
A detailed description of the instrument and its performance is given in \citet{Weisskopf2022}.

The data have been processed with the publicly available {\sc ixpeobssim} package\footnote{\url{https://github.com/lucabaldini/ixpeobssim}} version 30.2.2 \citep{Baldini2022} using CalDB released on 2022 November 17. 
Source photons were collected in a circular region with radius $R_{\rm src}=60\arcsec$ centered on the \xper position. 
Following recommendations from \cite{Di_Marco2023} for bright sources, the background was not subtracted from the data, because it appears to be negligible ($\sim$1\% of the total signal). 
The event arrival times were corrected to the Solar system barycenter using the standard {\tt barycorr} tool from the {\sc ftools} package and accounting for the effects of binary motion using the orbital parameters by \cite{2001ApJ...546..455D}. 
The resulting light curve of \xper in the 2--8\,keV band is shown in the upper panel of Fig.\,\ref{fig:ixpe-lc}. Because the source does not demonstrate significant difference in the average count rate in the two observational segments, the subsequent scientific analysis was performed using the joint dataset.

The same extraction procedure was applied to all three Stokes parameters $I$, $Q$ and $U$.
In order to use the $\chi^2$ statistics, the flux (Stokes parameter $I$) energy spectra were rebinned to have at least 30 counts per energy channel. 
The energy binning obtained for the Stokes parameter $I$ was also applied to the spectra of the Stokes parameters $Q$ and $U$. 
The subsequent spectral fitting was performed with the {\sc xspec} package \citep{Arn96} using the instrument response functions of version 10. 
Taking into account the high count statistic and negligible background level, the unweighted approach has been applied. 
The uncertainties are given at the 68.3 per cent confidence level for a single parameter of interest unless stated otherwise.

\subsection{\textit{SRG}/ART-XC}

To control the temporal and spectral properties of X Persei in the broader energy band, the quasi-simultaneous observations  with the {\it Spectrum-Roentgen-Gamma observatory} ({\it SRG}, \citealt{2021A&A...656A.132S}) Mikhail Pavlinsky ART-XC telescope \citep{2021A&A...650A..42P} were carried out on MJD~59813 with an $\sim$84~ks net exposure. The ART-XC telescope includes seven independent modules and provides imaging, timing and spectroscopy in the 4--30 keV energy range. ART-XC data were processed with the analysis software \textsc{artproducts} v1.0 and the CALDB  version 20220908. The corresponding light curves of \xper in the 4--24\,keV band is shown in the lower panels of  Fig.\,\ref{fig:ixpe-lc}.

\begin{figure}
\centering
\includegraphics[width=0.95\linewidth]{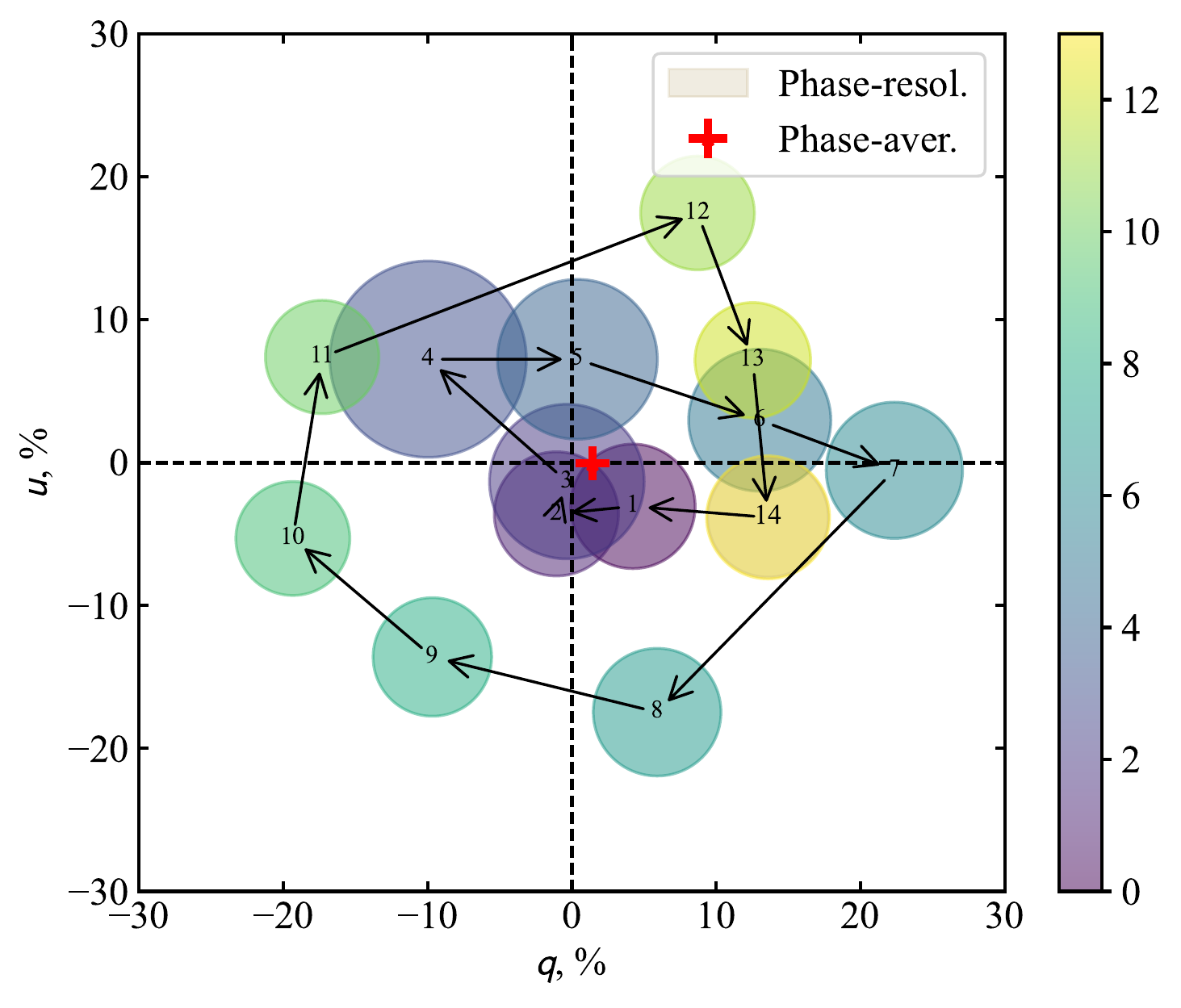}
\caption{Variations of the phase-resolved normalized Stokes parameters $q=Q/I$ and $u=U/I$ (from using \texttt{pcube}) over the spin phase of \xper in the 3--8~keV energy band averaged over three DUs. Colour coding and the size of the corresponding circle correspond to the phase bin number and $1\sigma$ uncertainty, respectively.
The phase-averaged value is shown with the red cross. }
 \label{fig:pcub}
\end{figure}

\begin{figure*}
\centering
\includegraphics[width=0.313\linewidth]{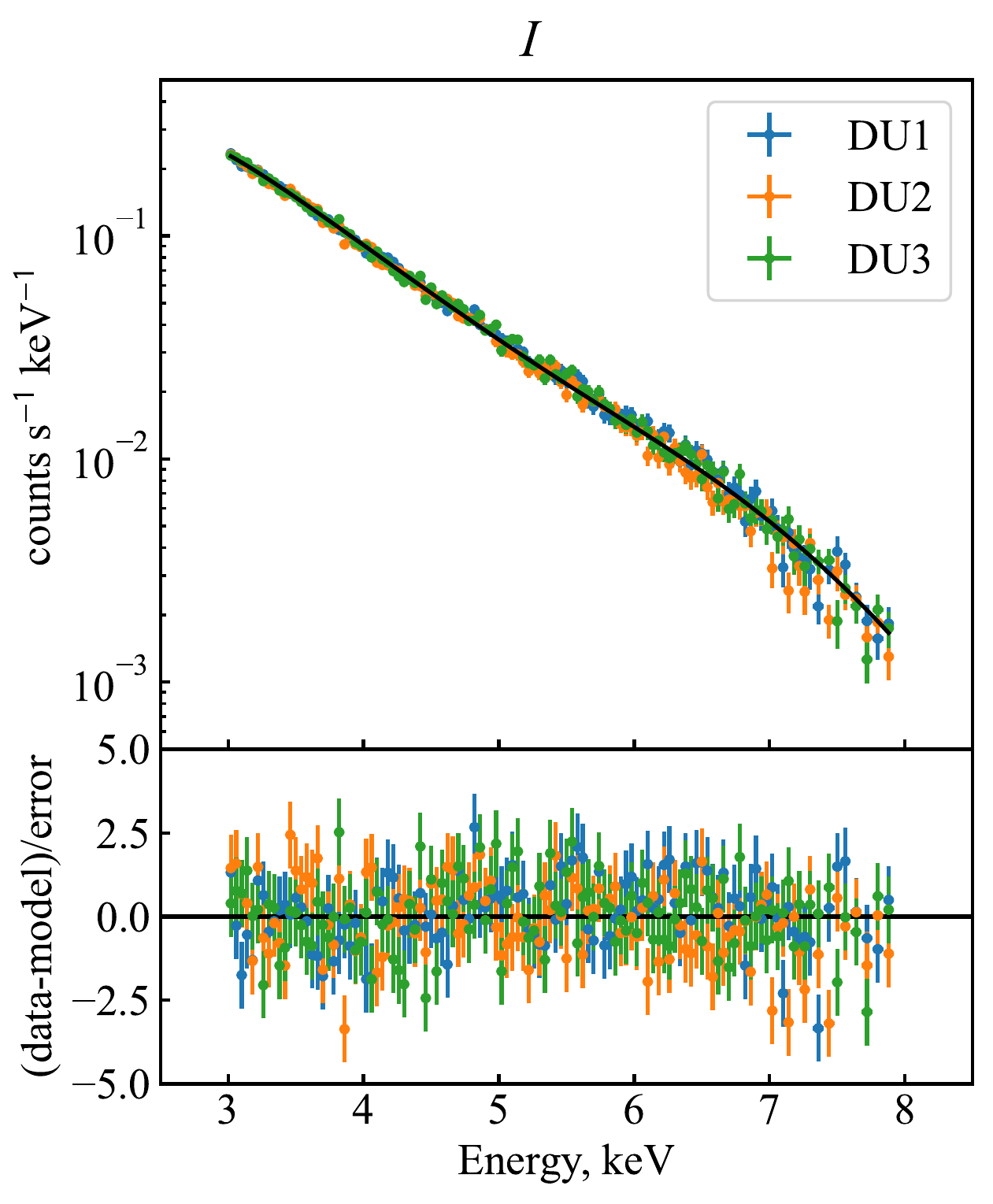}
\includegraphics[width=0.32\linewidth]{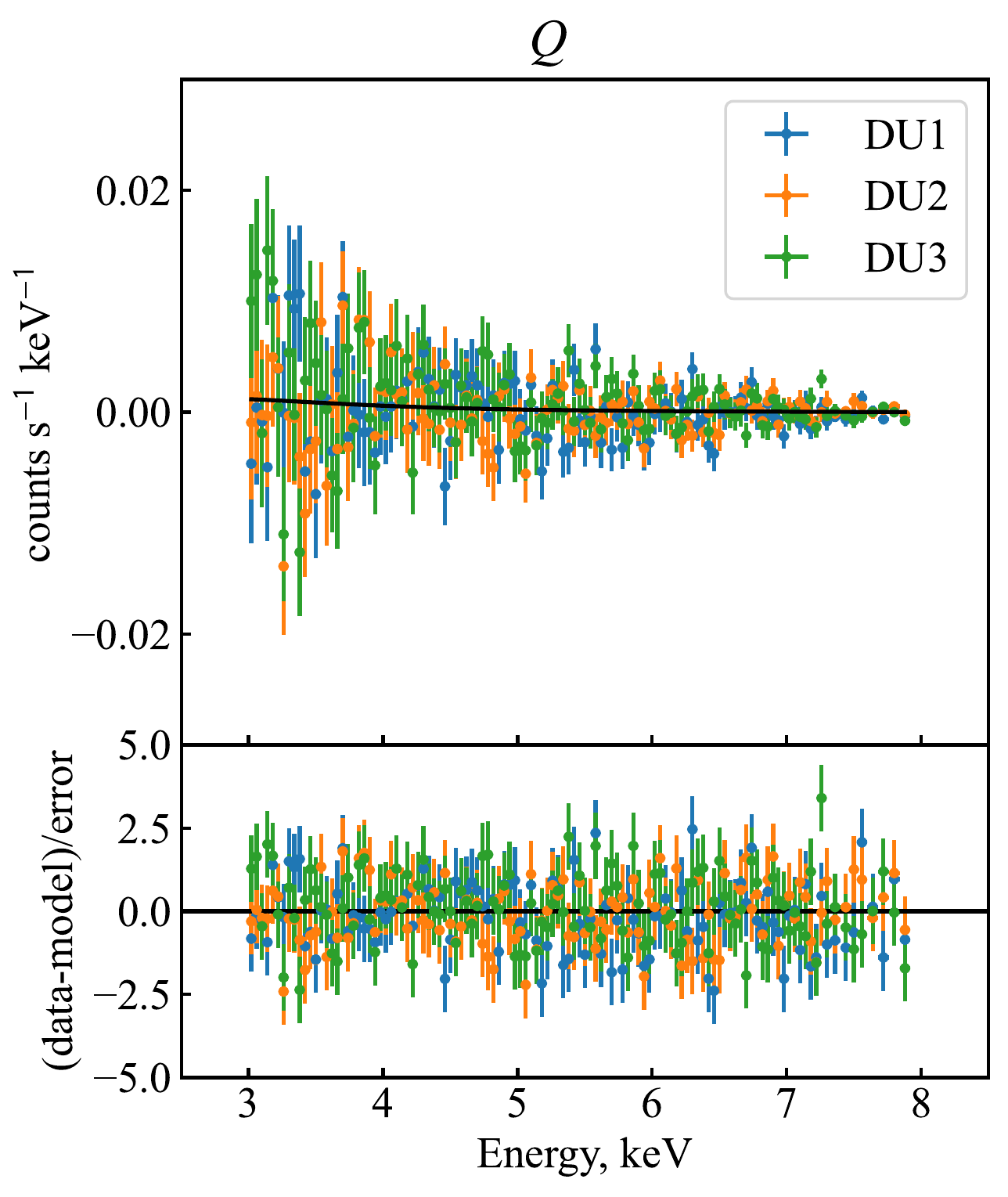}
\includegraphics[width=0.32\linewidth]{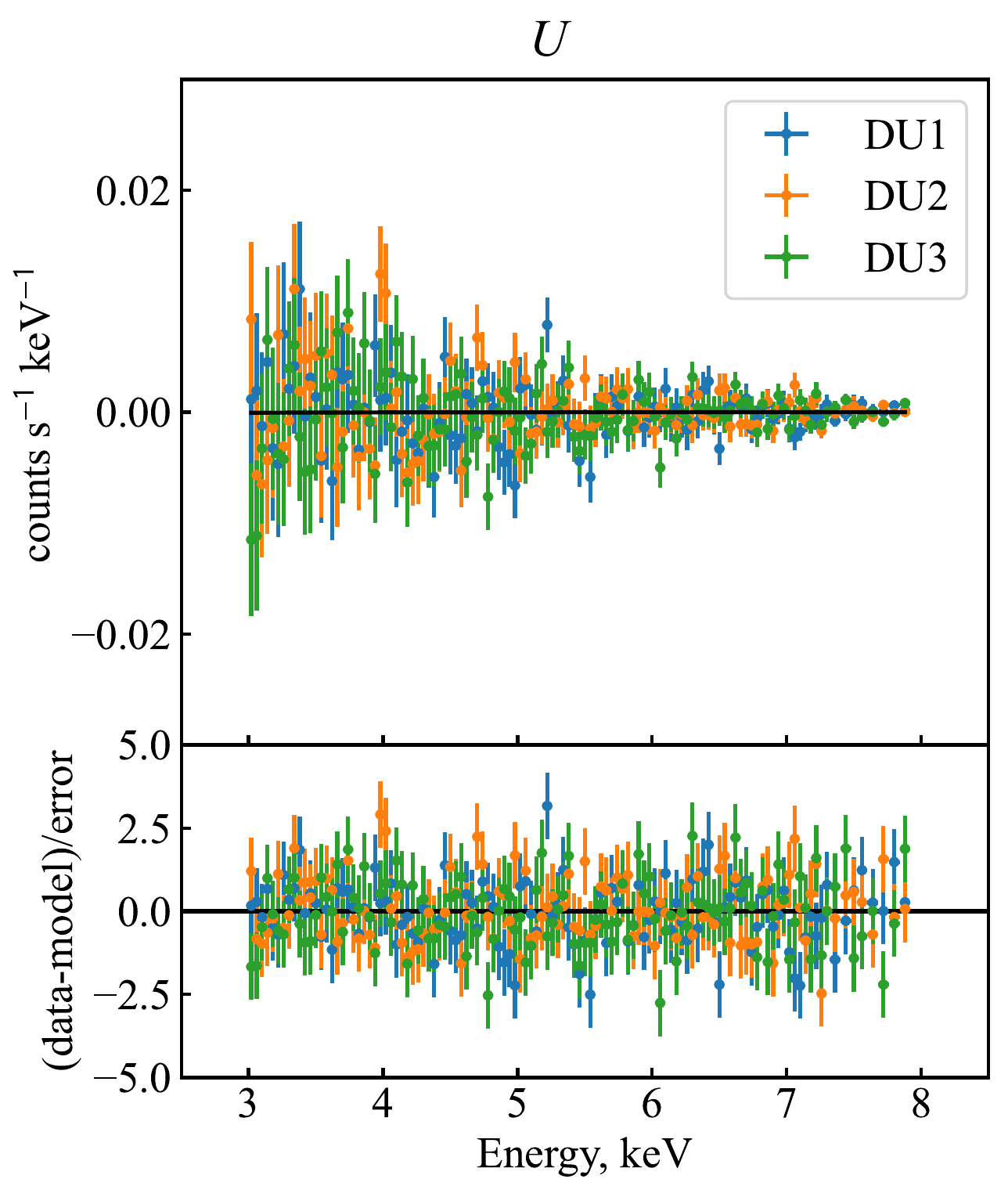}
\caption{Energy distributions of the Stokes parameters $I$, $Q$ and $U$ averaged over the spin phase of \xper obtained with \ixpe in the 3--8~keV band. The best-fitting model is shown with the black solid line. Bottom panels of each plot represent the residuals between the data and the model normalized for the error. Data from the three \ixpe detectors are shown with different colours: DU1 in blue, DU2 in orange and DU3 in green.  
}
 \label{fig:spec}
\end{figure*}

\section{Results}
\label{sec:Results}

Considering that the main goal of our investigation is pulse-phase resolved spectro-polarimetric analysis, we first determined pulsation period for the period covered by \ixpe data. The pulse period was first approximately determined as 833.14\,s using the event data (with event arrival times corrected for effects of orbital motion of the satellite and neutron star as described above) and $Z^2$ statistics \citep{1983A&A...128..245B}. This value was then refined using the phase-connection technique \citep{1981ApJ...247.1003D} to $P_{\rm spin}=833.214(10)$~s (the uncertainty is reported at $1\sigma$ confidence level). The folded pulse profile in different energy bands based on the \ixpe and ART-XC data (see Fig.~\ref{fig:pprof}) demonstrates a typical shape for the source with only minor dependence on energy.

Following standard procedures for \ixpe data analysis 
of XRPs \citep{2022NatAs...6.1433D,2022ApJ...941L..14T}, we performed polarimetric analysis of the data using the formalism of \cite{2015APh....68...45K} implemented in the {\tt pcube} algorithm in the {\tt xpbin} tool as a part of the {\sc ixpeobssim} package \citep{Baldini2022} as well as using spectro-polarimetric analysis with the the {\sc xspec} package \citep{2017ApJ...838...72S}. 
At the first step we explored the energy dependence of the polarimetric properties of \xper. 
We found that polarization is undetectable below 3 keV in the pulse phase-averaged as well as in the phase-resolved data. 
At the same time, above 3 keV the PD and the polarization angle (PA, measured from north to east) can be significantly measured and are consistent within the uncertainties (see Sect.~\ref{sec:geom} for the energy dependence and a possible reason for a low PD below 3~keV). 
Therefore, we excluded the 2--3~keV data from the following analysis.

Pulse-phase averaged polarimetric analysis using the {\tt pcube} algorithm in the 3--8 keV band resulted in a very low, consistent with zero, polarization with the normalized Stokes parameters of $q=Q/I=1.4\pm1.2\%$ and $u=U/I=0.0\pm1.2\%$ (see red cross in Fig.\,\ref{fig:pcub}). 
However, similarly to the others XRPs studied with \ixpe, phase-resolved analysis revealed a very strong variability of $q$ and $u$ over the pulse phase (Fig.\,\ref{fig:pcub}). 
We see that the PD$=\sqrt{q^2+u^2}$ varies from being consistent with zero to $\sim 20$ per cent.
The non-detection in the phase-averaged data can be thus attributed to a rotation of the PA with the pulse phase.

At the next step, we performed spectro-polarimetric analysis with \textsc{xspec} to account for the energy dispersion and the spectral shape. We fitted  jointly the $I$, $Q$ and $U$ spectra prepared with the PHA1, PHA1Q, and PHA1U algorithms in the {\tt xpbin} tool. To avoid problems with the mismatch in spectral calibrations of \ixpe and ART-XC telescopes and the fact that observations only partly overlap, for the following analysis only data from \ixpe were used.

\begin{table}
    \caption{Spectral parameters of the best-fitting model for the phase-averaged \ixpe data on \xper in the 3--8~keV band using \textsc{xspec}.}
    \label{tab:spec-aver}
    \centering
    \begin{tabular}{rll}
    \hline\hline
    Parameter & Value & Units\\ \hline
    $N_{\rm H}$  & 0.15$^{\rm fixed}$ &  $10^{22}$~cm$^{-2}$ \\
    const$_{\rm DU2}$ & 0.95$\pm0.01$ & \\
    const$_{\rm DU3}$ & 0.91$\pm0.01$ & \\
    Photon index & $1.72\pm0.02$ & \\
    PD & 1.8$\pm1.0$ & \% \\
    PA & $-$1.3$\pm16.0$ & deg \\
    Flux (2--8~keV) & 2.76$\pm0.04$& $10^{-10}$ erg~cm$^{-2}$~s$^{-1}$ \\
    Luminosity (2--8~keV) & $1.2\times10^{34}$  & erg~s$^{-1}$ at $d=600$~pc\\ 
    $\chi^{2}$ (d.o.f.) & 1044 (1038) & \\
    \hline
    \end{tabular}
 \begin{flushleft}{
 \textit{Note.} 
Quoted uncertainties are at the 68.3 per cent confidence level. Systematic uncertainty of 5\% was added in order to obtain the reliable errors on the model parameters.
}\end{flushleft} 
\end{table}

The spectrum of \xper is relatively simple and below 10 keV can be well described with an absorbed power law. The interstellar absorption was accounted for using the \texttt{tbabs} model with abundances adopted from \citet{Wilms2000}. Moreover, the fit appeared to be insensitive to the $N_{\rm H}$ value and it was fixed at $0.15\times10^{22}$~cm$^{-2}$ \citep[see e.g.][]{1998ApJ...509..897D}.
The PD and PA values were derived from the \texttt{polconst} model assuming their independence of the energy. 
The available statistic does not allow us to investigate the energy dependence in the phase-resolved data.
A cross-calibration multiplicative constant (\texttt{const} in {\sc xspec}) was introduced to the model in order to take into account possible discrepancies in the DU's effective areas (with DU1 constant fixed to unity). The quality of the phase-averaged spectrum approximation can be seen in Fig.~\ref{fig:spec} with the best-fitting parameters presented in Table~\ref{tab:spec-aver}.

\begin{figure}
\centering
\includegraphics[width=0.90\linewidth]{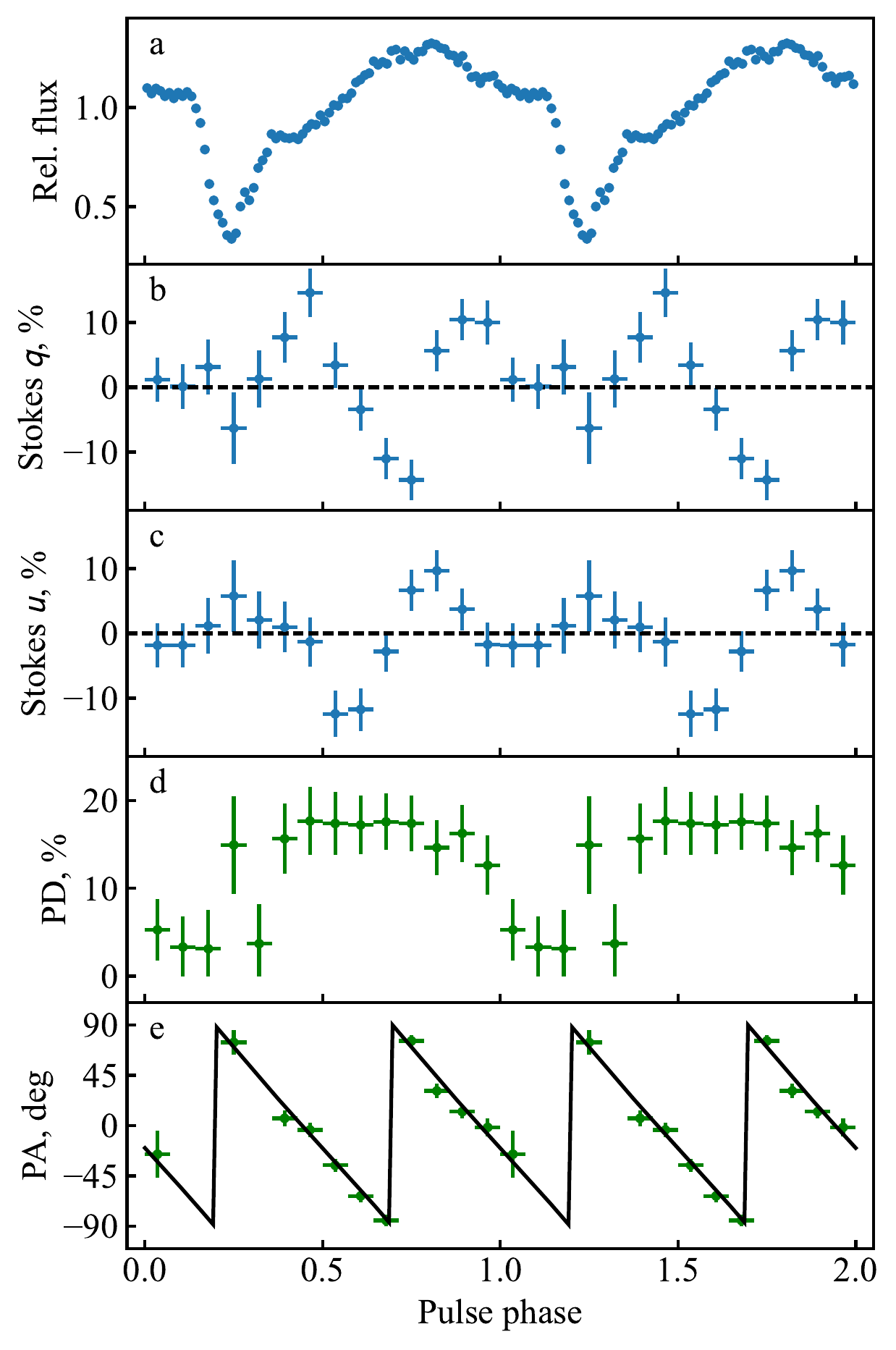}
\caption{Dependence of the relative
flux (\textit{panel a}), normalized  $q=Q/I$ and $u=U/I$ Stokes parameters (\textit{panels b} and \textit{c}), PD and PA (\textit{panels d} and \textit{e}) on the pulse phase in the 3--8 keV energy band. Data from the three \ixpe telescopes are combined. 
Data for three phase bins are missing the PA values because they are not constrained. 
The black solid line in the bottom panel corresponds to the best-fitting RVM with the following parameters: $i_{\rm p}=162\degr$, $\theta=90\degr$, $\chi_{\rm p}=70\degr$, and $\phi_0/2\pi=0.75$. 
 } 
 \label{fig:ixpe-st.pd.pa}
\end{figure}

\begin{table*}
\caption{Spectral parameters for the phase-resolved spectro-polarimetric analysis of the \xper data with \textsc{xspec}.  }
    \centering
    \begin{tabular}{cccccc}
    \hline\hline
 \#   &   Phase  &     Photon index &  PD & PA   & $\chi^{2}$ (d.o.f.) \\ 
    &      &      &  (\%) & (deg)  &   \\
          \hline
1 & 0.000--0.071 &  1.64$\pm0.04$  & 5.3$\pm3.5$  & $-$25.5$\pm21.0$  & 563 (569) \\
2 & 0.071--0.143 &  1.73$\pm0.04$  & 3.3$\pm3.3$  & \dots & 522 (550) \\
3 & 0.143--0.214 &  1.69$\pm0.04$  & 3.1$^{+4.4}_{-3.1}$  & \dots & 501 (460) \\
4 & 0.214--0.286 &  1.37$\pm0.04$  & 14.9$\pm5.6$  &  74.4$\pm11.0$  & 345 (362) \\
5 & 0.286--0.357 &  1.79$\pm0.04$  & 3.7$^{+4.5}_{-3.7}$  & \dots & 435 (449) \\
6 & 0.357--0.429 &  1.70$\pm0.04$  & 15.6$\pm4.0$  &  6.5$\pm7.4$  & 536 (488) \\
7 & 0.429--0.500 &  1.58$\pm0.04$  & 17.6$\pm3.9$  &  3.9$\pm6.2$  & 545 (517) \\
8 & 0.500--0.571 &  1.60$\pm0.04$  & 17.4$\pm3.6$  & $-$35.5$\pm6.0$  & 524 (544) \\
9 & 0.571--0.643 &  1.67$\pm0.04$  & 17.2$\pm3.3$  & $-$62.9$\pm5.6$  & 667 (597) \\
10 & 0.643--0.714 &  1.63$\pm0.04$  & 17.6$\pm3.2$  & $-$84.8$\pm5.3$  & 590 (611) \\
11 & 0.714--0.786 &  1.56$\pm0.04$  & 17.4$\pm3.2$  &  75.7$\pm5.2$  & 624 (612) \\
12 & 0.786--0.857 &  1.59$\pm0.04$  & 14.6$\pm3.1$  &  31.0$\pm6.2$  & 666 (620) \\
13 & 0.857--0.929 &  1.67$\pm0.04$  & 16.3$\pm3.3$  &  12.4$\pm5.7$  & 622 (602) \\
14 & 0.929--1.000 &  1.59$\pm0.04$  & 12.6$\pm3.4$  &  $-1.6\pm7.8$  & 532 (581) \\
    \hline
    \end{tabular}
 \begin{flushleft}{
 \textit{Note.} 
The cross-normalization constants as well as the $N_{\rm H}$ value are frozen at the values derived from the phase-averaged analysis. 
The missing values of the PA correspond to non-detection of the polarization.  
}\end{flushleft} 
   \label{tab:fit_phbin}
\end{table*}

\begin{figure*}
\centering
\includegraphics[width=0.9\linewidth]{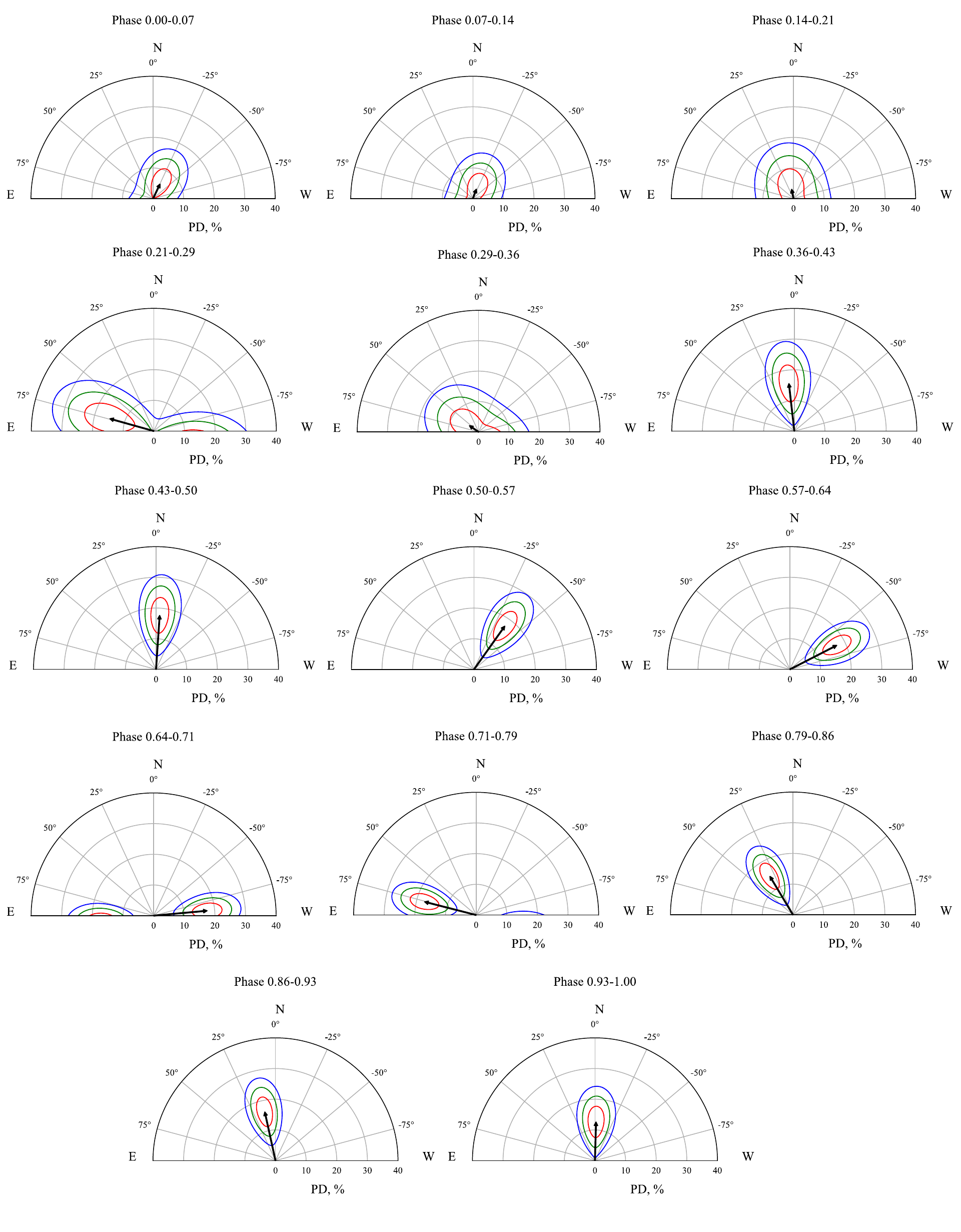}
\caption{Polarization vector of \xper as a function of the pulse phase based on the spectro-polarimetric analysis. The PD and PA at 68.27, 95.45 and 99.73 per cent  confidence levels (in red, green and blue colour, respectively) are shown for 14 phase intervals. 
}
\label{fig:cont-resol}
\end{figure*}

The same spectral model was applied to the phase-resolved data. All the spectral parameters were allowed to vary, except the cross-calibration constants, which were fixed to values obtained from the phase-averaged fit. The obtained results are summarized in Figs~\ref{fig:ixpe-st.pd.pa}, \ref{fig:cont-resol} and Table~\ref{tab:fit_phbin}. 
The PD is correlated with the pulsed flux, similarly to what was recently found in another XRP with luminosity much below the critical value, \gro \citep{Tsygankov23}. The PA varies from $-$90\degr\ to +90\degr, covering the widest range among the XRPs observed until now with \ixpe. During the pulse period, the PA makes two complete revolutions.

\section{Discussion}
\label{sec:disc}

\subsection{Atmospheric structure and polarization mechanism}

We estimate the bolometric accretion luminosity of \xper to be  below $\sim 5\times10^{34}\,\ergs$.
For the surface magnetic field of about $10^{13}\,{\rm G}$, this luminosity is well below the critical value $L_{\rm crit}\approx 3\times 10^{37}\,\ergs$ \citep{1976MNRAS.175..395B,2015MNRAS.447.1847M}, above which the radiative force becomes high enough to stop the accretion flow above the NS surface.
Under the condition of $L\ll L_{\rm crit}$, the influence of radiation pressure on the dynamics of the accretion flow is negligible and the accretion results in a hot spot geometry of the emission region. 
The observed two-component structure of the broadband X-ray energy spectrum and the luminosity of \xper indicate that the X-rays are emitted from the atmosphere under the conditions where the upper optically thin layer are overheated by the deceleration of accreting gas \citep{1969SvA....13..175Z,2018A&A...619A.114S}.
In this case, the low-energy component of the X-ray spectrum is produced by the thermal emission of underlying cold atmospheric layers, while the high-energy component is shaped by the cyclotron emission and subsequent multiple Compton resonant scatterings in the hot electron gas \citep{2021MNRAS.503.5193M,2021A&A...651A..12S}.

It is expected that a NS atmosphere with an inverse temperature profile produces a pencil beam emission pattern, which is suppressed along the NS surface normal: the hot upper layers contribute more to the X-ray energy flux leaving the NS atmosphere at larger angles with respect to the local normal.
In the case of a hot spot geometry, the PD below the cyclotron resonance is expected to be higher for photons leaving the atmosphere at larger angles to the local magnetic field direction because of a stronger dependence of scattering and absorption cross-sections on  polarization  \citep{1988ApJ...324.1056M}. 
Thus, the expected correlation between the flux and the PD is in agreement with the observations (compare panels a and d in Fig.\,\ref{fig:ixpe-st.pd.pa}). 
It means that the minimal X-ray energy flux and PD correspond to the smallest angle between NS magnetic axis and line of sight.
A similar correlation between the variations of X-ray energy flux and PD was detected recently in the XRP \gro, where strongly magnetized accreting NS was also observed  below the critical luminosity \citep{Tsygankov23}.

The low polarization degree observed in \xper is probably related to the inverse temperature profile in NS the atmosphere.
In this case, two effects come into play.
On one hand, the upper atmosphere in local thermodynamic equilibrium emits many photons for which the absorption and scattering cross-sections are larger. 
Because the emission occurs in an optically thin upper layer, these photons freely leave the atmosphere and equalize the difference between the fluxes in the two polarization modes \citep{2021MNRAS.503.5193M}.
In this scenario, the difference between fluxes in the two modes tend to be smaller at lower energies (see Fig.\,5 in \citealt{2021MNRAS.503.5193M}).
On the other hand, the estimates show that the temperature between the upper overheated layer and the lower cold atmosphere changes rapidly and there is a region with a large temperature gradient \citep{2018A&A...619A.114S}. 
Under the condition of hydrostatic equilibrium, the temperature jump is associated with a jump in mass density and a large density range is present at the boundary between the cold atmosphere and the overheated upper layer. 
If the vacuum resonance density
\beq 
\rho_{\rm V}\approx 10^{-4}\,B_{12}^2 E_{\rm keV}^2\,\,{\rm g\,cm^{-3}},
\eeq
where $B_{12}$ is the field strength in units of $10^{12}\,{\rm G}$ and $E_{\rm keV}$ is the photon energy in keV \citep{2003ApJ...588..962L},
falls into this density range, a polarization mode conversion is expected at the lower boundary of the overheated upper atmospheric layer.
It has  been shown previously that mode conversion in an overheated upper atmospheric layer  leads to flux equalisation in the two polarization modes and a corresponding reduction in the linear PD (see Supplementary Materials in \citealt{2022NatAs...6.1433D}).

\begin{figure*}
\centering
\includegraphics[width=0.75\linewidth]{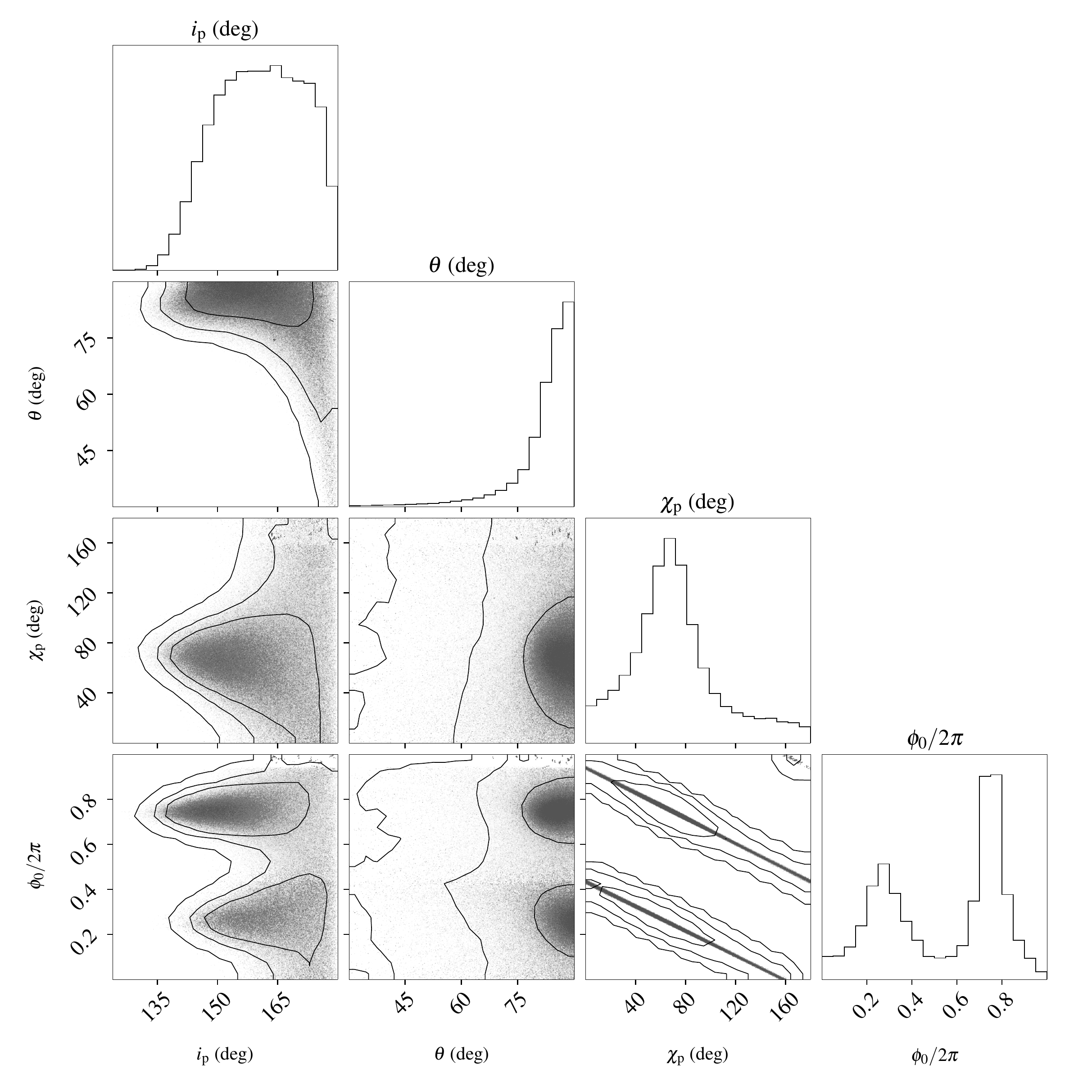}
\caption{Corner plot of the posterior distribution for the RVM parameters using the PA dependence on the spin phase. The two-dimensional contours correspond to 68.27, 95.45 and 99.73 per cent  confidence levels. 
} 
 \label{fig:rvm}
\end{figure*}

\subsection{Constraints on the geometry}
\label{sec:geom}

The PA makes two complete turns during the pulsation period (see Fig.\,\ref{fig:pcub} and  Fig.\,\ref{fig:ixpe-st.pd.pa}e). 
Polarization of X-ray photons with respect to the local direction of the magnetic field is conserved in the NS magnetosphere within the adiabatic radius 
$R_{\rm ad} \approx 7.6\times 10^6\,B_{12}^{0.4}E_{\rm keV}^{0.2}$\,cm \citep{Heyl00,2016MNRAS.459.3585G}. 
Thus, it is expected that the PA follows the projection of the magnetic field axis in the sky (or is perpendicular to this projection, depending on the polarization composition of the X-rays leaving the NS atmosphere). 
A full turn of the magnetic axis projection during the pulse period becomes possible under the condition 
\beq\label{eq:condition} 
i_{\rm p}<\theta,
\eeq
where $i_{\rm p}$ is the pulsar spin inclination angle and $\theta$ is the magnetic obliquity, i.e. the angle between the rotation and the NS magnetic field axis. 
Interaction of the accretion flow with NS magnetosphere leads to the alignment of the NS spin with the orbit on a time scale  \citep{2021MNRAS.505.1775B}
\beq 
\tau \approx 600\,I_{45}\left(\frac{\dot{M}}{\dot{M}_{\rm Edd}}\right)^{-6/7}
\left(\frac{M}{M_\odot}\right)^{-9/7}\mu_{30}^{-2/7}\,\,{\rm yr},
\eeq 
where  $I_{45}$ is the NS moment of inertia in units of $10^{45}\,{\rm g\,cm^2}$, $M_{\rm Edd}\approx 1.6\times 10^{18}\,{\rm g\,s^{-1}}$ is the Eddington mass accretion rate onto a NS, and $\mu_{30}$ is the NS magnetic moment in units of $10^{30}\,{\rm G\,cm^3}$.
For the observed mass accretion rate in \xper and under the assumption of a surface magnetic field $B\sim 10^{13}\,{\rm G}$, we get the time scale required for the alignment $\tau\sim 4\times 10^5\,{\rm yr}$.
The estimated time scale of spin axis alignment is much smaller than the age reported for the optical companion in \xper by \citet{1997MNRAS.286..549L}: $\sim 5\times 10^6\,{\rm yr}$ and sufficiently small to assume that the magnetic field of the NS has not experienced a significant decay (see \citealt{2019LRCA....5....3P} for review and Fig.\,20 there).
Thus, the rotational axis of the NS in \xper is likely aligned with the orbital axis of the binary system. 
In this case, one would expect $i_{\rm p}\sim i_{\rm orb}\approx 30\degr$ and the condition (\ref{eq:condition}) can be easily satisfied.

To determine the geometry of \xper, we follow the procedure applied earlier to Her~X-1 \citep{2022NatAs...6.1433D}, Cen~X-3 \citep{2022ApJ...941L..14T} and \gro \citep{Tsygankov23} and fit the rotating vector model \citep[RVM,][]{1969ApL.....3..225R,Poutanen20RVM},
\begin{equation} \label{eq:pa_rvm}
\tan (\mbox{PA}\!-\!\chi_{\rm p})\!=\! \frac{-\sin \theta\ \sin (\phi-\phi_0)}
{\sin i_{\rm p} \cos \theta\!  - \! \cos i_{\rm p} \sin \theta  \cos (\phi\!-\!\phi_0) } ,
\end{equation} 
to the observed variation of the PA with phase $\phi$. 
Here $\chi_{\rm p}$ is the position angle of the pulsar's rotation axis (assuming that radiation escapes predominantly in the O-mode).
We use the affine invariant Markov chain Monte Carlo ensemble sampler {\sc emcee} package of {\sc python} \citep{2013PASP..125..306F}. 
All the parameters in equation~(\ref{eq:pa_rvm}) were left free.  
Because polarization is sensitive to the sense of rotation, the prior for the pulsar inclination $i_{\rm p}$ is $[0\degr,180\degr]$ with the probability density $\propto\sin i_{\rm p}$.
The inclination is constrained at $i_{\rm p}=162\degr\pm12\degr$ at a 68 per cent  confidence level and $i_{\rm p}>135\degr$ at a 99 per cent  confidence level. 
This is consistent with the orbital inclination $i_{\rm orb}\in [26\degr,33\degr]$ reported by \citet{2001ApJ...546..455D} using the X-ray pulse arrival times and measured mass function (note that these constraints are equivalent to $i_{\rm orb}\approx150\degr$ as the sense of rotation cannot be determined from such data). 
The probability distribution of the position angle $\chi_{\rm p}$ has a broad peak at $\chi_{\rm p,O}\approx70\degr$ with a 1$\sigma$ width of about 30\degr.  
We note here that there is an alternative solution $\chi_{\rm p}=\chi_{\rm p,O}+180\degr=250\degr$, because only the orientation of the polarization plane can be measured. 
Furthermore, if radiation is dominated by the X-mode, then the pulsar spin orientation is $\chi_{\rm p}=\chi_{\rm p,X}=\chi_{\rm p,O}\pm90\degr$.

The spin phase $\phi_0$ when the closest pole is directed towards the observer has two maxima at $\phi_0/(2\pi)\approx0.25$ and 0.75, because the two rotations of the PA are shifted by nearly exactly half a period.
Finally, the magnetic obliquity tends to values close to $90\degr$, with the lower limit being $75\degr$ at a 68 per cent confidence level.  
However, an orthogonal magnetic dipole in \xper (i.e. $\theta\approx 90\degr$),  contradicts the behaviour of the normalized Stokes parameters $q$ and $u$ during the pulse period (see Fig.\,\ref{fig:pcub}) and a proper analysis of the data requires assumptions on the beam patterns in both polarization modes.

A sharp dip observed in the pulse profile at phase $\sim 0.25$ (Fig.\,\ref{fig:ixpe-st.pd.pa}), when the normal to one of the poles is close to the line of sight, can be related to an eclipse of the hot spot by the accretion stream above the NS surface.
However, the mass accretion rate in \xper is very low and the optical thickness of the accretion stream above NS surface seems to be too small to support this hypothesis.
Yet, the suggestion of a spot eclipse at this phase is also supported by observations of a sharp spike of the X-ray flux in the centre of the dip seen, for example, in the \textit{Suzaku} data \citep{2017MNRAS.470..713M}
with the corresponding increase of the hardness ratio already observed with \textit{EXOSAT} \citep{1989ApJ...346..469R} and \textit{Ginga} \citep{1996ApJ...472..341R}, which implies that here we are looking directly along the magnetic dipole axis.

Once the geometry of the pulsar is obtained from the RVM, we can now combine all the data to  investigate the energy dependence of the polarimetric properties of \xper in the phase-averaged data.  
Here we used all the data in the 2--8 keV \ixpe band.
To eliminate the effect of the PA rotation over pulse, the PAs were frozen at the predictions of the best-fitting RVM at a given phase (see Fig.~\ref{fig:ixpe-st.pd.pa}e). 
We then performed a joint spectro-polarimetric fit of the data collected in all 14 phases where instead of \texttt{polconst} model we use the \texttt{pollin} model $\mbox{PD}(E)=A_1+A_{\rm slope}(E-1)$, where $E$ is the photon energy in keV.
Photon indices were fixed at the values presented in Table~\ref{tab:fit_phbin} and the PD parameters were tied for all phases. 
As the result we obtained the phase-averaged PD at 1 keV of $A_1=-6.3\%\pm2.0\%$ and slope of $A_{\rm slope}=5.1\%\pm0.7\%$~keV$^{-1}$. 
Such a dependence corresponds to the zero PD at $\sim$2.25~keV (which explains the non-detection of polarization in the 2--3 keV band) and around 30\% at 8~keV. 
The negative value of the PD is equivalent to the rotation of the PA by 90\degr\ for photons below 2~keV relative to those above 3~keV. 
Using the F-test we estimated the significance of the improvement of the fit with the \texttt{pollin} model compared to \texttt{polconst} as $2\times10^{-14}$.
It is worth mentioning that a similar behaviour of the PD increasing with energy was recently discovered in another XRP Vela~X-1 \citep{Forsblom23}, where the PD crosses zero around 3.5~keV.

\section{Summary}
\label{sec:sum}

The results of polarimetric studies of XRP \xper can be summarized as follows: 
\begin{enumerate}[leftmargin=12pt]
\item 
Linear polarization was not detected in the phase-averaged date from \xper.
However, the phase-resolved polarimetry revealed variable PD and PA. 
The energy-averaged PD in the 3--8~keV band reaches $\sim$20 per cent, while the PA makes two complete revolutions during the pulsation period.  
The PD is detected to be positively correlated with variations of the X-ray energy flux during the observed pulse period (see panels a and d in Fig.\,\ref{fig:ixpe-st.pd.pa}). 
A positive correlation between the flux and the PD was already reported in the low-luminosity XRP \gro \citep{Tsygankov23}, and probably is a typical feature for XRPs at a luminosity well below the critical one.
\item 
The observed variations of the PA, making two full turns per spin period, become possible under the condition that the magnetic obliquity is larger than the inclination of the NS rotation axis. 
Application of the RVM to the data gives the following set of geometrical parameters: 
the inclination is constrained $i_{\rm p}=162\degr\pm12\degr$ ($>135\degr$) at 68 (99) per cent confidence level, consistent with the orbital inclination reported in \xper. 
The position angle has been determined as $70\degr\pm30\degr$ or $250\degr\pm30\degr$ (assuming that the observed photons were polarized in the O-mode), with alternative solutions rotated by 90\degr, if radiation escapes in the X-mode.
The magnetic obliquity has a preference towards $90\degr$ with the lower limit being $75\degr$ at a 68 per cent  confidence level. 
High magnetic obliquity potentially makes \xper the second discovered orthogonal rotator after \gro. 
\item
Eliminating the effect of rotation of the PA with the pulse phase applying the predictions of the  best-fitting RVM, a strong energy dependence of the PD was discovered with the PD increasing from 0\% at $\sim$2~keV to 30\% at 8~keV similarly to another XRP Vela~X-1 \citep{Forsblom23}. 
\item  
Similarity of the determined NS inclination and the orbital inclination supports the hypothesis of a strong NS magnetic field in \xper, which would make the relaxation period of the NS towards spin equilibrium and alignment with the orbital axis relatively fast even under conditions of a rather low mass accretion rate. 
\item
The low PD and its correlation with the X-ray flux during the pulse period is consistent with the expected inverse temperature profile in the NS atmosphere at low mass accretion rates. 
Under this condition, the beam pattern of X-ray radiation leaving the NS surface is expected to be suppressed along the normal to the surface. 
Thus, a smaller X-ray flux within the pulse period can be related to a smaller angle between the line of sight and the magnetic dipole. 
\end{enumerate}

\section*{Acknowledgements}

The Imaging X-ray Polarimetry Explorer (IXPE) is a joint US and Italian mission.  
The US contribution is supported by the National Aeronautics and Space Administration (NASA) and led and managed by its Marshall Space Flight Center (MSFC), with industry partner Ball Aerospace (contract NNM15AA18C).  
The Italian contribution is supported by the Italian Space Agency (Agenzia Spaziale Italiana, ASI) through contract ASI-OHBI-2017-12-I.0, agreements ASI-INAF-2017-12-H0 and ASI-INFN-2017.13-H0, and its Space Science Data Center (SSDC) with agreements ASI-INAF-2022-14-HH.0 and ASI-INFN 2021-43-HH.0, and by the Istituto Nazionale di Astrofisica (INAF) and the Istituto Nazionale di Fisica Nucleare (INFN) in Italy.
This research used data products provided by the IXPE Team (MSFC, SSDC, INAF, and INFN) and distributed with additional software tools by the High-Energy Astrophysics Science Archive Research Center (HEASARC), at NASA Goddard Space Flight Center (GSFC). 

We acknowledge support from UKRI Stephen Hawking fellowship and the Netherlands Organization for Scientific Research Veni fellowship  (AAM), the Academy of Finland grants 333112, 349144, 349373, and 349906 (SST, JP), the V\"ais\"al\"a Foundation (SST), the German Academic Exchange Service (DAAD) travel grant 57525212 (VD, VFS), the German Research Foundation (DFG) grant \mbox{WE 1312/53-1} (VFS), the Russian Science Foundation grant 19-12-00423 (AS, AAL, IAM, ANS, AES), and the CNES fellowship grant (DG-C).

\section*{Data availability}

The \ixpe data used in this paper are publicly available in the HEASARC database.


\begin{thebibliography}{}
\makeatletter
\relax
\def\mn@urlcharsother{\let\do\@makeother \do\$\do\&\do\#\do\^\do\_\do\%\do\~}
\def\mn@doi{\begingroup\mn@urlcharsother \@ifnextchar [ {\mn@doi@}
  {\mn@doi@[]}}
\def\mn@doi@[#1]#2{\def\@tempa{#1}\ifx\@tempa\@empty \href
  {http://dx.doi.org/#2} {doi:#2}\else \href {http://dx.doi.org/#2} {#1}\fi
  \endgroup}
\def\mn@eprint#1#2{\mn@eprint@#1:#2::\@nil}
\def\mn@eprint@arXiv#1{\href {http://arxiv.org/abs/#1} {{\tt arXiv:#1}}}
\def\mn@eprint@dblp#1{\href {http://dblp.uni-trier.de/rec/bibtex/#1.xml}
  {dblp:#1}}
\def\mn@eprint@#1:#2:#3:#4\@nil{\def\@tempa {#1}\def\@tempb {#2}\def\@tempc
  {#3}\ifx \@tempc \@empty \let \@tempc \@tempb \let \@tempb \@tempa \fi \ifx
  \@tempb \@empty \def\@tempb {arXiv}\fi \@ifundefined
  {mn@eprint@\@tempb}{\@tempb:\@tempc}{\expandafter \expandafter \csname
  mn@eprint@\@tempb\endcsname \expandafter{\@tempc}}}

\bibitem[\protect\citeauthoryear{{Arnaud}}{{Arnaud}}{1996}]{Arn96}
{Arnaud} K.~A.,  1996, in {Jacoby} G.~H.,  {Barnes} J.,  eds,  ASP Conf. Ser.
  Vol. 101, Astronomical Data Analysis Software and Systems V. Astron. Soc.
  Pac., San Francisco, pp 17--20

\bibitem[\protect\citeauthoryear{{Bailer-Jones}, {Rybizki}, {Fouesneau},
  {Demleitner}  \& {Andrae}}{{Bailer-Jones} et~al.}{2021}]{2021AJ....161..147B}
{Bailer-Jones} C.~A.~L.,  {Rybizki} J.,  {Fouesneau} M.,  {Demleitner} M.,
  {Andrae} R.,  2021, \mn@doi [\aj] {10.3847/1538-3881/abd806}, \href
  {https://ui.adsabs.harvard.edu/abs/2021AJ....161..147B} {161, 147}

\bibitem[\protect\citeauthoryear{{Baldini} et~al.,}{{Baldini}
  et~al.}{2021}]{2021APh...13302628B}
{Baldini} L.,  et~al., 2021, \mn@doi [Astroparticle Physics]
  {10.1016/j.astropartphys.2021.102628}, \href
  {https://ui.adsabs.harvard.edu/abs/2021APh...13302628B} {133, 102628}

\bibitem[\protect\citeauthoryear{{Baldini} et~al.,}{{Baldini}
  et~al.}{2022}]{Baldini2022}
{Baldini} L.,  et~al., 2022, \mn@doi [SoftwareX] {10.1016/j.softx.2022.101194},
  \href {https://ui.adsabs.harvard.edu/abs/2022SoftX..1901194B} {19, 101194}

\bibitem[\protect\citeauthoryear{{Basko} \& {Sunyaev}}{{Basko} \&
  {Sunyaev}}{1976}]{1976MNRAS.175..395B}
{Basko} M.~M.,  {Sunyaev} R.~A.,  1976, \mn@doi [\mnras]
  {10.1093/mnras/175.2.395}, \href
  {https://ui.adsabs.harvard.edu/abs/1976MNRAS.175..395B} {175, 395}

\bibitem[\protect\citeauthoryear{{Biryukov} \& {Abolmasov}}{{Biryukov} \&
  {Abolmasov}}{2021}]{2021MNRAS.505.1775B}
{Biryukov} A.,  {Abolmasov} P.,  2021, \mn@doi [\mnras]
  {10.1093/mnras/stab1378}, \href
  {https://ui.adsabs.harvard.edu/abs/2021MNRAS.505.1775B} {505, 1775}

\bibitem[\protect\citeauthoryear{{Buccheri} et~al.,}{{Buccheri}
  et~al.}{1983}]{1983A&A...128..245B}
{Buccheri} R.,  et~al., 1983, \aap, \href
  {https://ui.adsabs.harvard.edu/abs/1983A&A...128..245B} {128, 245}

\bibitem[\protect\citeauthoryear{{Caiazzo} \& {Heyl}}{{Caiazzo} \&
  {Heyl}}{2021}]{Caiazzo21}
{Caiazzo} I.,  {Heyl} J.,  2021, \mn@doi [\mnras] {10.1093/mnras/staa3428},
  \href {https://ui.adsabs.harvard.edu/abs/2021MNRAS.501..109C} {501, 109}

\bibitem[\protect\citeauthoryear{{Coburn}, {Heindl}, {Gruber}, {Rothschild},
  {Staubert}, {Wilms}  \& {Kreykenbohm}}{{Coburn}
  et~al.}{2001}]{2001ApJ...552..738C}
{Coburn} W.,  {Heindl} W.~A.,  {Gruber} D.~E.,  {Rothschild} R.~E.,  {Staubert}
  R.,  {Wilms} J.,   {Kreykenbohm} I.,  2001, \mn@doi [\apj] {10.1086/320565},
  \href {https://ui.adsabs.harvard.edu/abs/2001ApJ...552..738C} {552, 738}

\bibitem[\protect\citeauthoryear{{Deeter}, {Boynton}  \& {Pravdo}}{{Deeter}
  et~al.}{1981}]{1981ApJ...247.1003D}
{Deeter} J.~E.,  {Boynton} P.~E.,   {Pravdo} S.~H.,  1981, \mn@doi [\apj]
  {10.1086/159110}, \href
  {https://ui.adsabs.harvard.edu/abs/1981ApJ...247.1003D} {247, 1003}

\bibitem[\protect\citeauthoryear{{Delgado-Mart{\'\i}}, {Levine}, {Pfahl}  \&
  {Rappaport}}{{Delgado-Mart{\'\i}} et~al.}{2001}]{2001ApJ...546..455D}
{Delgado-Mart{\'\i}} H.,  {Levine} A.~M.,  {Pfahl} E.,   {Rappaport} S.~A.,
  2001, \mn@doi [\apj] {10.1086/318236}, \href
  {https://ui.adsabs.harvard.edu/abs/2001ApJ...546..455D} {546, 455}

\bibitem[\protect\citeauthoryear{{Di Marco} et~al.,}{{Di Marco}
  et~al.}{2023}]{Di_Marco2023}
{Di Marco} A.,  et~al., 2023, \mn@doi [\aj] {10.3847/1538-3881/acba0f}, \href
  {https://ui.adsabs.harvard.edu/abs/2023arXiv230202927D} {165, 143}

\bibitem[\protect\citeauthoryear{{Di Salvo}, {Burderi}, {Robba}  \&
  {Guainazzi}}{{Di Salvo} et~al.}{1998}]{1998ApJ...509..897D}
{Di Salvo} T.,  {Burderi} L.,  {Robba} N.~R.,   {Guainazzi} M.,  1998, \mn@doi
  [\apj] {10.1086/306525}, \href
  {https://ui.adsabs.harvard.edu/abs/1998ApJ...509..897D} {509, 897}

\bibitem[\protect\citeauthoryear{{Doroshenko}, {Santangelo}, {Kreykenbohm}  \&
  {Doroshenko}}{{Doroshenko} et~al.}{2012}]{2012A&A...540L...1D}
{Doroshenko} V.,  {Santangelo} A.,  {Kreykenbohm} I.,   {Doroshenko} R.,  2012,
  \mn@doi [\aap] {10.1051/0004-6361/201218878}, \href
  {https://ui.adsabs.harvard.edu/abs/2012A&A...540L...1D} {540, L1}

\bibitem[\protect\citeauthoryear{{Doroshenko} et~al.,}{{Doroshenko}
  et~al.}{2022}]{2022NatAs...6.1433D}
{Doroshenko} V.,  et~al., 2022, \mn@doi [Nature Astronomy]
  {10.1038/s41550-022-01799-5}, \href
  {https://ui.adsabs.harvard.edu/abs/2022NatAs...6.1433D} {6, 1433}

\bibitem[\protect\citeauthoryear{{Foreman-Mackey}, {Hogg}, {Lang}  \&
  {Goodman}}{{Foreman-Mackey} et~al.}{2013}]{2013PASP..125..306F}
{Foreman-Mackey} D.,  {Hogg} D.~W.,  {Lang} D.,   {Goodman} J.,  2013, \mn@doi
  [\pasp] {10.1086/670067}, \href
  {https://ui.adsabs.harvard.edu/abs/2013PASP..125..306F} {125, 306}

\bibitem[\protect\citeauthoryear{{Forsblom} et~al.,}{{Forsblom}
  et~al.}{2023}]{Forsblom23}
{Forsblom} S.~V.,  et~al., 2023, \apjl, in press, \href
  {https://ui.adsabs.harvard.edu/abs/2023arXiv230301800F} {p. arXiv:2303.01800}

\bibitem[\protect\citeauthoryear{{Gnedin} \& {Pavlov}}{{Gnedin} \&
  {Pavlov}}{1974}]{1974JETP...38..903G}
{Gnedin} Y.~N.,  {Pavlov} G.~G.,  1974, Soviet Journal of Experimental and
  Theoretical Physics, \href
  {https://ui.adsabs.harvard.edu/abs/1974JETP...38..903G} {38, 903}

\bibitem[\protect\citeauthoryear{{Gonz{\'a}lez Caniulef}, {Zane}, {Taverna},
  {Turolla}  \& {Wu}}{{Gonz{\'a}lez Caniulef}
  et~al.}{2016}]{2016MNRAS.459.3585G}
{Gonz{\'a}lez Caniulef} D.,  {Zane} S.,  {Taverna} R.,  {Turolla} R.,   {Wu}
  K.,  2016, \mn@doi [\mnras] {10.1093/mnras/stw804}, \href
  {https://ui.adsabs.harvard.edu/abs/2016MNRAS.459.3585G} {459, 3585}

\bibitem[\protect\citeauthoryear{{Harding} \& {Lai}}{{Harding} \&
  {Lai}}{2006}]{2006RPPh...69.2631H}
{Harding} A.~K.,  {Lai} D.,  2006, \mn@doi [Reports on Progress in Physics]
  {10.1088/0034-4885/69/9/R03}, \href
  {https://ui.adsabs.harvard.edu/abs/2006RPPh...69.2631H} {69, 2631}

\bibitem[\protect\citeauthoryear{{Heyl} \& {Shaviv}}{{Heyl} \&
  {Shaviv}}{2000}]{Heyl00}
{Heyl} J.~S.,  {Shaviv} N.~J.,  2000, \mn@doi [\mnras]
  {10.1046/j.1365-8711.2000.03076.x}, \href
  {https://ui.adsabs.harvard.edu/abs/2000MNRAS.311..555H} {311, 555}

\bibitem[\protect\citeauthoryear{{Kaminker}, {Pavlov}  \&
  {Shibanov}}{{Kaminker} et~al.}{1983}]{1983Ap&SS..91..167K}
{Kaminker} A.~D.,  {Pavlov} G.~G.,   {Shibanov} I.~A.,  1983, \mn@doi [\apss]
  {10.1007/BF00650222}, \href
  {https://ui.adsabs.harvard.edu/abs/1983Ap&SS..91..167K} {91, 167}

\bibitem[\protect\citeauthoryear{{Kislat}, {Clark}, {Beilicke}  \&
  {Krawczynski}}{{Kislat} et~al.}{2015}]{2015APh....68...45K}
{Kislat} F.,  {Clark} B.,  {Beilicke} M.,   {Krawczynski} H.,  2015, \mn@doi
  [Astroparticle Physics] {10.1016/j.astropartphys.2015.02.007}, \href
  {https://ui.adsabs.harvard.edu/abs/2015APh....68...45K} {68, 45}

\bibitem[\protect\citeauthoryear{{Kong} et~al.,}{{Kong}
  et~al.}{2022}]{2022ApJ...933L...3K}
{Kong} L.-D.,  et~al., 2022, \mn@doi [\apjl] {10.3847/2041-8213/ac7711}, \href
  {https://ui.adsabs.harvard.edu/abs/2022ApJ...933L...3K} {933, L3}

\bibitem[\protect\citeauthoryear{{Lai} \& {Ho}}{{Lai} \&
  {Ho}}{2003}]{2003ApJ...588..962L}
{Lai} D.,  {Ho} W. C.~G.,  2003, \mn@doi [\apj] {10.1086/374334}, \href
  {https://ui.adsabs.harvard.edu/abs/2003ApJ...588..962L} {588, 962}

\bibitem[\protect\citeauthoryear{{Lutovinov}, {Tsygankov}  \&
  {Chernyakova}}{{Lutovinov} et~al.}{2012}]{2012MNRAS.423.1978L}
{Lutovinov} A.,  {Tsygankov} S.,   {Chernyakova} M.,  2012, \mn@doi [\mnras]
  {10.1111/j.1365-2966.2012.21036.x}, \href
  {https://ui.adsabs.harvard.edu/abs/2012MNRAS.423.1978L} {423, 1978}

\bibitem[\protect\citeauthoryear{{Lyubimkov}, {Rostopchin}, {Roche}  \&
  {Tarasov}}{{Lyubimkov} et~al.}{1997}]{1997MNRAS.286..549L}
{Lyubimkov} L.~S.,  {Rostopchin} S.~I.,  {Roche} P.,   {Tarasov} A.~E.,  1997,
  \mn@doi [\mnras] {10.1093/mnras/286.3.549}, \href
  {https://ui.adsabs.harvard.edu/abs/1997MNRAS.286..549L} {286, 549}

\bibitem[\protect\citeauthoryear{{Maitra}, {Raichur}, {Pradhan}  \&
  {Paul}}{{Maitra} et~al.}{2017}]{2017MNRAS.470..713M}
{Maitra} C.,  {Raichur} H.,  {Pradhan} P.,   {Paul} B.,  2017, \mn@doi [\mnras]
  {10.1093/mnras/stx1281}, \href
  {https://ui.adsabs.harvard.edu/abs/2017MNRAS.470..713M} {470, 713}

\bibitem[\protect\citeauthoryear{{Meszaros}}{{Meszaros}}{1992}]{1992herm.book.....M}
{Meszaros} P.,  1992, {High-energy radiation from magnetized neutron stars}.
University of Chicago Press, Chicago

\bibitem[\protect\citeauthoryear{{Meszaros} \& {Nagel}}{{Meszaros} \&
  {Nagel}}{1985a}]{1985ApJ...298..147M}
{Meszaros} P.,  {Nagel} W.,  1985a, \mn@doi [\apj] {10.1086/163594}, \href
  {https://ui.adsabs.harvard.edu/abs/1985ApJ...298..147M} {298, 147}

\bibitem[\protect\citeauthoryear{{Meszaros} \& {Nagel}}{{Meszaros} \&
  {Nagel}}{1985b}]{1985ApJ...299..138M}
{Meszaros} P.,  {Nagel} W.,  1985b, \mn@doi [\apj] {10.1086/163687}, \href
  {https://ui.adsabs.harvard.edu/abs/1985ApJ...299..138M} {299, 138}

\bibitem[\protect\citeauthoryear{{Meszaros}, {Novick}, {Szentgyorgyi}, {Chanan}
   \& {Weisskopf}}{{Meszaros} et~al.}{1988}]{1988ApJ...324.1056M}
{Meszaros} P.,  {Novick} R.,  {Szentgyorgyi} A.,  {Chanan} G.~A.,   {Weisskopf}
  M.~C.,  1988, \mn@doi [\apj] {10.1086/165962}, \href
  {https://ui.adsabs.harvard.edu/abs/1988ApJ...324.1056M} {324, 1056}

\bibitem[\protect\citeauthoryear{{Mushtukov} \& {Tsygankov}}{{Mushtukov} \&
  {Tsygankov}}{2022}]{2022arXiv220414185M}
{Mushtukov} A.,  {Tsygankov} S.,  2022, arXiv e-prints, \href
  {https://ui.adsabs.harvard.edu/abs/2022arXiv220414185M} {p. arXiv:2204.14185}

\bibitem[\protect\citeauthoryear{{Mushtukov}, {Suleimanov}, {Tsygankov}  \&
  {Poutanen}}{{Mushtukov} et~al.}{2015}]{2015MNRAS.447.1847M}
{Mushtukov} A.~A.,  {Suleimanov} V.~F.,  {Tsygankov} S.~S.,   {Poutanen} J.,
  2015, \mn@doi [\mnras] {10.1093/mnras/stu2484}, \href
  {https://ui.adsabs.harvard.edu/abs/2015MNRAS.447.1847M} {447, 1847}

\bibitem[\protect\citeauthoryear{{Mushtukov}, {Suleimanov}, {Tsygankov}  \&
  {Portegies Zwart}}{{Mushtukov} et~al.}{2021}]{2021MNRAS.503.5193M}
{Mushtukov} A.~A.,  {Suleimanov} V.~F.,  {Tsygankov} S.~S.,   {Portegies Zwart}
  S.,  2021, \mn@doi [\mnras] {10.1093/mnras/stab811}, \href
  {https://ui.adsabs.harvard.edu/abs/2021MNRAS.503.5193M} {503, 5193}

\bibitem[\protect\citeauthoryear{{Nakajima}, {Negoro}, {Mihara}, {Sugizaki},
  {Yatabe}  \& {Makishima}}{{Nakajima} et~al.}{2019}]{2019IAUS..346..131N}
{Nakajima} M.,  {Negoro} H.,  {Mihara} T.,  {Sugizaki} M.,  {Yatabe} F.,
  {Makishima} K.,  2019, \mn@doi [IAU Symposium] {10.1017/S1743921319001820},
  \href {https://ui.adsabs.harvard.edu/abs/2019IAUS..346..131N} {346, 131}

\bibitem[\protect\citeauthoryear{{Pavlinsky} et~al.,}{{Pavlinsky}
  et~al.}{2021}]{2021A&A...650A..42P}
{Pavlinsky} M.,  et~al., 2021, \mn@doi [\aap] {10.1051/0004-6361/202040265},
  \href {https://ui.adsabs.harvard.edu/abs/2021A&A...650A..42P} {650, A42}

\bibitem[\protect\citeauthoryear{{Pons} \& {Vigan{\`o}}}{{Pons} \&
  {Vigan{\`o}}}{2019}]{2019LRCA....5....3P}
{Pons} J.~A.,  {Vigan{\`o}} D.,  2019, \mn@doi [Living Reviews in Computational
  Astrophysics] {10.1007/s41115-019-0006-7}, \href
  {https://ui.adsabs.harvard.edu/abs/2019LRCA....5....3P} {5, 3}

\bibitem[\protect\citeauthoryear{{Poutanen}}{{Poutanen}}{2020}]{Poutanen20RVM}
{Poutanen} J.,  2020, \mn@doi [\aap] {10.1051/0004-6361/202038689}, \href
  {https://ui.adsabs.harvard.edu/abs/2020A&A...641A.166P} {641, A166}

\bibitem[\protect\citeauthoryear{{Radhakrishnan} \& {Cooke}}{{Radhakrishnan} \&
  {Cooke}}{1969}]{1969ApL.....3..225R}
{Radhakrishnan} V.,  {Cooke} D.~J.,  1969, \aplett, \href
  {https://ui.adsabs.harvard.edu/abs/1969ApL.....3..225R} {3, 225}

\bibitem[\protect\citeauthoryear{{Reig} \& {Roche}}{{Reig} \&
  {Roche}}{1999}]{1999MNRAS.306..100R}
{Reig} P.,  {Roche} P.,  1999, \mn@doi [\mnras]
  {10.1046/j.1365-8711.1999.02473.x}, \href
  {https://ui.adsabs.harvard.edu/abs/1999MNRAS.306..100R} {306, 100}

\bibitem[\protect\citeauthoryear{{Robba} \& {Warwick}}{{Robba} \&
  {Warwick}}{1989}]{1989ApJ...346..469R}
{Robba} N.~R.,  {Warwick} R.~S.,  1989, \mn@doi [\apj] {10.1086/168027}, \href
  {https://ui.adsabs.harvard.edu/abs/1989ApJ...346..469R} {346, 469}

\bibitem[\protect\citeauthoryear{{Robba}, {Burderi}, {Wynn}, {Warwick}  \&
  {Murakami}}{{Robba} et~al.}{1996}]{1996ApJ...472..341R}
{Robba} N.~R.,  {Burderi} L.,  {Wynn} G.~A.,  {Warwick} R.~S.,   {Murakami} T.,
   1996, \mn@doi [\apj] {10.1086/178067}, \href
  {https://ui.adsabs.harvard.edu/abs/1996ApJ...472..341R} {472, 341}

\bibitem[\protect\citeauthoryear{{Soffitta} et~al.,}{{Soffitta}
  et~al.}{2021}]{2021AJ....162..208S}
{Soffitta} P.,  et~al., 2021, \mn@doi [\aj] {10.3847/1538-3881/ac19b0}, \href
  {https://ui.adsabs.harvard.edu/abs/2021AJ....162..208S} {162, 208}

\bibitem[\protect\citeauthoryear{{Sokolova-Lapa} et~al.,}{{Sokolova-Lapa}
  et~al.}{2021}]{2021A&A...651A..12S}
{Sokolova-Lapa} E.,  et~al., 2021, \mn@doi [\aap]
  {10.1051/0004-6361/202040228}, \href
  {https://ui.adsabs.harvard.edu/abs/2021A&A...651A..12S} {651, A12}

\bibitem[\protect\citeauthoryear{{Staubert} et~al.,}{{Staubert}
  et~al.}{2019}]{2019A&A...622A..61S}
{Staubert} R.,  et~al., 2019, \mn@doi [\aap] {10.1051/0004-6361/201834479},
  \href {https://ui.adsabs.harvard.edu/abs/2019A&A...622A..61S} {622, A61}

\bibitem[\protect\citeauthoryear{{Strohmayer}}{{Strohmayer}}{2017}]{2017ApJ...838...72S}
{Strohmayer} T.~E.,  2017, \mn@doi [\apj] {10.3847/1538-4357/aa643d}, \href
  {https://ui.adsabs.harvard.edu/abs/2017ApJ...838...72S} {838, 72}

\bibitem[\protect\citeauthoryear{{Suleimanov}, {Poutanen}  \&
  {Werner}}{{Suleimanov} et~al.}{2018}]{2018A&A...619A.114S}
{Suleimanov} V.~F.,  {Poutanen} J.,   {Werner} K.,  2018, \mn@doi [\aap]
  {10.1051/0004-6361/201833581}, \href
  {https://ui.adsabs.harvard.edu/abs/2018A&A...619A.114S} {619, A114}

\bibitem[\protect\citeauthoryear{{Suleimanov}, {Mushtukov}, {Ognev},
  {Doroshenko}  \& {Werner}}{{Suleimanov} et~al.}{2022}]{2022MNRAS.517.4022S}
{Suleimanov} V.~F.,  {Mushtukov} A.~A.,  {Ognev} I.,  {Doroshenko} V.~A.,
  {Werner} K.,  2022, \mn@doi [\mnras] {10.1093/mnras/stac2935}, \href
  {https://ui.adsabs.harvard.edu/abs/2022MNRAS.517.4022S} {517, 4022}

\bibitem[\protect\citeauthoryear{{Sunyaev} et~al.,}{{Sunyaev}
  et~al.}{2021}]{2021A&A...656A.132S}
{Sunyaev} R.,  et~al., 2021, \mn@doi [\aap] {10.1051/0004-6361/202141179},
  \href {https://ui.adsabs.harvard.edu/abs/2021A&A...656A.132S} {656, A132}

\bibitem[\protect\citeauthoryear{{Tsygankov}, {Mushtukov}, {Suleimanov},
  {Doroshenko}, {Abolmasov}, {Lutovinov}  \& {Poutanen}}{{Tsygankov}
  et~al.}{2017}]{2017A&A...608A..17T}
{Tsygankov} S.~S.,  {Mushtukov} A.~A.,  {Suleimanov} V.~F.,  {Doroshenko} V.,
  {Abolmasov} P.~K.,  {Lutovinov} A.~A.,   {Poutanen} J.,  2017, \mn@doi [\aap]
  {10.1051/0004-6361/201630248}, \href
  {https://ui.adsabs.harvard.edu/abs/2017A&A...608A..17T} {608, A17}

\bibitem[\protect\citeauthoryear{{Tsygankov}, {Rouco Escorial}, {Suleimanov},
  {Mushtukov}, {Doroshenko}, {Lutovinov}, {Wijnands}  \&
  {Poutanen}}{{Tsygankov} et~al.}{2019a}]{2019MNRAS.483L.144T}
{Tsygankov} S.~S.,  {Rouco Escorial} A.,  {Suleimanov} V.~F.,  {Mushtukov}
  A.~A.,  {Doroshenko} V.,  {Lutovinov} A.~A.,  {Wijnands} R.,   {Poutanen} J.,
   2019a, \mn@doi [\mnras] {10.1093/mnrasl/sly236}, \href
  {https://ui.adsabs.harvard.edu/abs/2019MNRAS.483L.144T} {483, L144}

\bibitem[\protect\citeauthoryear{{Tsygankov}, {Doroshenko}, {Mushtukov},
  {Suleimanov}, {Lutovinov}  \& {Poutanen}}{{Tsygankov}
  et~al.}{2019b}]{2019MNRAS.487L..30T}
{Tsygankov} S.~S.,  {Doroshenko} V.,  {Mushtukov} A.~A.,  {Suleimanov} V.~F.,
  {Lutovinov} A.~A.,   {Poutanen} J.,  2019b, \mn@doi [\mnras]
  {10.1093/mnrasl/slz079}, \href
  {https://ui.adsabs.harvard.edu/abs/2019MNRAS.487L..30T} {487, L30}

\bibitem[\protect\citeauthoryear{{Tsygankov} et~al.,}{{Tsygankov}
  et~al.}{2022}]{2022ApJ...941L..14T}
{Tsygankov} S.~S.,  et~al., 2022, \mn@doi [\apjl] {10.3847/2041-8213/aca486},
  \href {https://ui.adsabs.harvard.edu/abs/2022ApJ...941L..14T} {941, L14}

\bibitem[\protect\citeauthoryear{{Tsygankov} et~al.,}{{Tsygankov}
  et~al.}{2023}]{Tsygankov23}
{Tsygankov} S.~S.,  et~al., 2023, \mn@doi [\aap, submitted]
  {10.48550/arXiv.2302.06680}, \href
  {https://ui.adsabs.harvard.edu/abs/2023arXiv230206680T} {p. arXiv:2302.06680}

\bibitem[\protect\citeauthoryear{{Wang} \& {Frank}}{{Wang} \&
  {Frank}}{1981}]{1981A&A....93..255W}
{Wang} Y.~M.,  {Frank} J.,  1981, \aap, \href
  {https://ui.adsabs.harvard.edu/abs/1981A&A....93..255W} {93, 255}

\bibitem[\protect\citeauthoryear{{Weisskopf} et~al.,}{{Weisskopf}
  et~al.}{2022}]{Weisskopf2022}
{Weisskopf} M.~C.,  et~al., 2022, \mn@doi [J. Astron. Telesc. Instrum. Syst.]
  {10.1117/1.JATIS.8.2.026002}, \href
  {https://ui.adsabs.harvard.edu/abs/2021arXiv211201269W} {8, 026002}

\bibitem[\protect\citeauthoryear{{White}, {Mason}, {Sanford}  \&
  {Murdin}}{{White} et~al.}{1976}]{1976MNRAS.176..201W}
{White} N.~E.,  {Mason} K.~O.,  {Sanford} P.~W.,   {Murdin} P.,  1976, \mn@doi
  [\mnras] {10.1093/mnras/176.1.201}, \href
  {https://ui.adsabs.harvard.edu/abs/1976MNRAS.176..201W} {176, 201}

\bibitem[\protect\citeauthoryear{{Wilms}, {Allen}  \& {McCray}}{{Wilms}
  et~al.}{2000}]{Wilms2000}
{Wilms} J.,  {Allen} A.,   {McCray} R.,  2000, \mn@doi [\apj] {10.1086/317016},
  \href {https://ui.adsabs.harvard.edu/\#abs/2000ApJ...542..914W} {542, 914}

\bibitem[\protect\citeauthoryear{{Yatabe}, {Makishima}, {Mihara}, {Nakajima},
  {Sugizaki}, {Kitamoto}, {Yoshida}  \& {Takagi}}{{Yatabe}
  et~al.}{2018}]{2018PASJ...70...89Y}
{Yatabe} F.,  {Makishima} K.,  {Mihara} T.,  {Nakajima} M.,  {Sugizaki} M.,
  {Kitamoto} S.,  {Yoshida} Y.,   {Takagi} T.,  2018, \mn@doi [\pasj]
  {10.1093/pasj/psy088}, \href
  {https://ui.adsabs.harvard.edu/abs/2018PASJ...70...89Y} {70, 89}

\bibitem[\protect\citeauthoryear{{Zel'dovich} \& {Shakura}}{{Zel'dovich} \&
  {Shakura}}{1969}]{1969SvA....13..175Z}
{Zel'dovich} Y.~B.,  {Shakura} N.~I.,  1969, \sovast, \href
  {http://adsabs.harvard.edu/abs/1969SvA....13..175Z} {13, 175}

\makeatother
\end{thebibliography}

\vspace{1cm} 

\noindent
\textit{
$^{1}$Astrophysics, Department of Physics, University of Oxford, Denys Wilkinson Building, Keble Road, Oxford OX1 3RH, UK \label{Oxford}\\
$^{2}$Leiden Observatory, Leiden University, NL-2300RA Leiden, The Netherlands \\
$^{3}$Department of Physics and Astronomy, FI-20014 University of Turku,  Finland \\
$^{4}$Institut f\"ur Astronomie und Astrophysik, Universit\"at T\"ubingen, Sand 1, D-72076 T\"ubingen, Germany \\
$^{5}$Department of Astronomy, Saint Petersburg State University, Saint-Petersburg 198504, Russia\\
$^{6}$Space Research Institute of the Russian Academy of Sciences, Profsoyuznaya Str. 84/32, Moscow 117997, Russia\\
$^{7}$INAF Istituto di Astrofisica e Planetologia Spaziali, Via del Fosso del Cavaliere 100, 00133 Roma, Italy \\
$^{8}$Department of Physics and Astronomy, University of British Columbia, Vancouver, BC V6T 1Z1, Canada\\
$^{9}$INAF Osservatorio Astronomico di Roma, Via Frascati 33, 00040 Monte Porzio Catone (RM), Italy \\
$^{10}$Institut de Recherche en Astrophysique et Plan\'etologie, UPS-OMP, CNRS, CNES, 9 avenue du Colonel Roche, BP 44346 31028, Toulouse CEDEX 4, France\\
$^{11}$International Space Science Institute, Hallerstrasse 6, 3012 Bern, Switzerland\\
$^{12}$Max Planck Institute for Astrophysics, Karl-Schwarzschild-Str 1, D-85741 Garching, Germany \\
$^{13}$Instituto de Astrof\'isica de Andaluc\'ia-CSIC, Glorieta de la Astronom\'ia s/n, 18008, Granada, Spain\\ 
$^{14}$INAF Osservatorio Astronomico di Roma, Via Frascati 33, 00078 Monte Porzio Catone (RM), Italy\\ 
$^{15}$Space Science Data Center, Agenzia Spaziale Italiana, Via del Politecnico snc, 00133 Roma, Italy\\ 
$^{16}$INAF -- Osservatorio Astronomico di Cagliari, via della Scienza 5, I-09047 Selargius (CA), Italy\\
$^{17}$Istituto Nazionale di Fisica Nucleare, Sezione di Pisa, Largo B. Pontecorvo 3, 56127 Pisa, Italy\\ 
$^{18}$Dipartimento di Fisica, Universit\`a di Pisa, Largo B. Pontecorvo 3, 56127 Pisa, Italy\\ 
$^{19}$NASA Marshall Space Flight Center, Huntsville, AL 35812, USA\\ 
$^{20}$Dipartimento di Matematica e Fisica, Universit\`a degli Studi Roma Tre, Via della Vasca Navale 84, I-00146 Roma, Italy\\
$^{21}$Istituto Nazionale di Fisica Nucleare, Sezione di Torino, Via Pietro Giuria 1, 10125 Torino, Italy\\ 
$^{22}$Dipartimento di Fisica,  Universit\`a degli Studi di Torino, Via Pietro Giuria 1, 10125 Torino, Italy\\ 
$^{23}$INAF Osservatorio Astrofisico di Arcetri, Largo Enrico Fermi 5, 50125 Firenze, Italy\\ 
$^{24}$Dipartimento di Fisica e Astronomia,  Universit\`a degli Studi di Firenze, Via Sansone 1, 50019 Sesto Fiorentino (FI), Italy\\ 
$^{25}$Istituto Nazionale di Fisica Nucleare, Sezione di Firenze, Via Sansone 1, 50019 Sesto Fiorentino (FI), Italy\\ 
$^{26}$ASI - Agenzia Spaziale Italiana, Via del Politecnico snc, 00133 Roma, Italy\\ 
$^{27}$Science and Technology Institute, Universities Space Research Association, Huntsville, AL 35805, USA \\
$^{28}$Istituto Nazionale di Fisica Nucleare, Sezione di Roma `Tor Vergata', Via della Ricerca Scientifica 1, 00133 Roma, Italy\\ 
$^{29}$Department of Physics and Kavli Institute for Particle Astrophysics and Cosmology, Stanford University, Stanford, California 94305, USA\\ 
$^{30}$Astronomical Institute of the Czech Academy of Sciences, Bo\v{c}n\'{i} II 1401/1, 14100 Praha 4, Czech Republic\\ 
$^{31}$RIKEN Cluster for Pioneering Research, 2-1 Hirosawa, Wako, Saitama 351-0198, Japan\\ 
$^{32}$California Institute of Technology, Pasadena, CA 91125, USA\\ 
$^{33}$Yamagata University,1-4-12 Kojirakawa-machi, Yamagata-shi 990-8560, Japan\\ 
$^{34}$Osaka University, 1-1 Yamadaoka, Suita, Osaka 565-0871, Japan\\
$^{35}$International Center for Hadron Astrophysics, Chiba University, Chiba 263-8522, Japan\\
$^{36}$Institute for Astrophysical Research, Boston University, 725 Commonwealth Avenue, Boston, MA 02215, USA\\ 
$^{37}$Department of Astrophysics, St. Petersburg State University, Universitetsky pr. 28, Petrodvoretz, 198504 St. Petersburg, Russia\\ 
$^{38}$University of New Hampshire, Department of Physics \& Astronomy and Space Science Center, 8 College Rd, Durham, NH 03824, USA\\
$^{39}$Physics Department and McDonnell Center for the Space Sciences, Washington University in St. Louis, St. Louis, MO 63130, USA\\ 
$^{40}$Finnish Centre for Astronomy with ESO,  20014 University of Turku, Finland\\ 
$^{41}$Istituto Nazionale di Fisica Nucleare, Sezione di Napoli, Strada Comunale Cinthia, 80126 Napoli, Italy\\
$^{42}$Universit\'{e} de Strasbourg, CNRS, Observatoire Astronomique de Strasbourg, UMR 7550, 67000 Strasbourg, France\\ 
$^{43}$MIT Kavli Institute for Astrophysics and Space Research, Massachusetts Institute of Technology, 77 Massachusetts Avenue, Cambridge, MA 02139, USA\\ 
$^{44}$Graduate School of Science, Division of Particle and Astrophysical Science, Nagoya University, Furo-cho, Chikusa-ku, Nagoya, Aichi 464-8602, Japan\\ 
$^{45}$Hiroshima Astrophysical Science Center, Hiroshima University, 1-3-1 Kagamiyama, Higashi-Hiroshima, Hiroshima 739-8526, Japan\\ 
$^{46}$University of Maryland, Baltimore County, Baltimore, MD 21250, USA\\
$^{47}$NASA Goddard Space Flight Center, Greenbelt, MD 20771, USA\\
$^{48}$Center for Research and Exploration in Space Science and Technology, NASA/GSFC, Greenbelt, MD 20771, USA\\
$^{49}$Department of Physics, The University of Hong Kong, Pokfulam, Hong Kong\\ 
$^{50}$Department of Astronomy and Astrophysics, Pennsylvania State University, University Park, PA 16802, USA\\ 
$^{51}$Universit\'{e} Grenoble Alpes, CNRS, IPAG, 38000 Grenoble, France\\ 
$^{52}$Center for Astrophysics, Harvard \& Smithsonian, 60 Garden St, Cambridge, MA 02138, USA\\ 
$^{53}$INAF Osservatorio Astronomico di Brera, Via E. Bianchi 46, 23807 Merate (LC), Italy\\ 
$^{54}$Dipartimento di Fisica e Astronomia, Universit\`a degli Studi di Padova, Via Marzolo 8, 35131 Padova, Italy\\ 
$^{55}$Dipartimento di Fisica, Universit\`a degli Studi di Roma `Tor Vergata', Via della Ricerca Scientifica 1, 00133 Roma, Italy\\ 
$^{56}$Department of Astronomy, University of Maryland, College Park, Maryland 20742, USA\\ 
$^{57}$Mullard Space Science Laboratory, University College London, Holmbury St Mary, Dorking, Surrey RH5 6NT, UK\\ 
$^{58}$Anton Pannekoek Institute for Astronomy \& GRAPPA, University of Amsterdam, Science Park 904, 1098 XH Amsterdam, The Netherlands\\ 
$^{59}$Guangxi Key Laboratory for Relativistic Astrophysics, School of Physical Science and Technology, Guangxi University, Nanning 530004, China\\ 
}

\bsp 
\label{lastpage}
\end{document}